\documentclass[aps,amsfonts,pra,showpacs,nobibnotes,nofootinbib,tightenlines]{revtex4}
\usepackage{graphicx}% Include figure files
\begin{document}
\newcommand{\beq}{\begin{equation}}
\newcommand{\eeq}{\end{equation}}
\newcommand{\barr}{\begin{eqnarray}}
\newcommand{\earr}{\end{eqnarray}}
\def\figwidth{7.5cm}
\newcommand{\rjdelj}[1]{[[ RJ June: #1 ]]}
\newcommand{\rjcommj}[1]{{\sc RJ June: #1}}
\newcommand{\rjaddj}[1]{{\bf RJ June: #1}}
\newcommand{\rjdelf}[1]{[[ RJ final: #1 ]]}
\newcommand{\rjcommf}[1]{{\sc RJ final: #1}}
\newcommand{\rjaddf}[1]{{\bf RJ final: #1}}
\newcommand{\ascomm}[1]{{\textsc{AS final: #1}}}
\newcommand{\asdrop}[1]{[[ AS - DELETED: #1 ]]}
\newcommand{\asadd}[1]{{\bf AS final: [[#1]]}}
\newcommand{\andy}[1]{ }
%% RLJ should de-comment the following line
%\newcommand{\asfigure}[3]{\BoxedEPSF{#1 scaled #2}}
%% RLJ should comment the following line
\newcommand{\asfigure}[3]{\includegraphics[width=#3]{#1}}
\def\cH{{\cal H}}
\def\cP{{\cal P}}
\def\cD{{\cal D}}
\def\cG{{\cal G}}
\def\cV{{\cal V}}
\def\cF{{\cal F}}
\def\cU{{\cal U}}
\def\cS{{\cal S}}
\def\cO{{\cal O}}
\def\cE{{\cal E}}
\def\bfA{{\bf A}}
\def\bfG{{\bf G}}
\def\bfn{{\bf n}}
\def\bfr{{\bf r}}
\def\bfV{{\bf V}}
\def\bft{{\bf t}}
\def\bfM{{\bf M}}
\def\bfP{{\bf P}}
\def\bra#1{\langle #1 |}
\def\ket#1{| #1 \rangle}
\newcommand{\p}{\partial}
%%%%%%%%%%%%%%%%%%%%%%%%new def%%%%%%%%%%%%%%%%%%%%%%%%%%%
%\newcommand{\ket}[1]{| #1 \rangle}
%\newcommand{\bra}[1]{\langle #1 |}
\def\coltwovector#1#2{\left({#1\atop#2}\right)}
\def\upp{\coltwovector10}
\def\downn{\coltwovector01}
\def\Ord#1{{\cal O}\left( #1\right)}
\def\bmp{\mbox{\boldmath $p$}}
\def\rhobar{\bar{\rho}}
\renewcommand{\Re}{{\rm Re}}
\renewcommand{\Im}{{\rm Im}}
\renewcommand{\theequation}{\thesection.\arabic{equation}}
%%%%%%%%%%%%%%%%%%%%%%%%%%%%%%%%%%%%%%%%%%%%%%%%%%%%%%%%%%
\def\ask{\marginpar{?? ask:  \hfill}}
\def\fin{\marginpar{fill in ... \hfill}}
\def\note{\marginpar{note \hfill}}
\def\check{\marginpar{check \hfill}}
\def\discuss{\marginpar{discuss \hfill}}
%%%%%%%%%%%%%%%%%%%%%%%%%%%%%%%%%
%Title of paper
\title{Casimir Effects: An Optical
Approach
\\
I.  Foundations and Examples
}\author{A.~Scardicchio}\email{scardicc@mit.edu}
\author{R.~L.~Jaffe}\email{jaffe@mit.edu}
\affiliation{Center for Theoretical Physics, \\ Laboratory for
   Nuclear Science and Department of Physics \\ Massachusetts
Institute
  of Technology \\ Cambridge, MA 02139, USA}
%\date{\today}
\begin{abstract}
    \noindent  We present the foundations of a new approach to
    the Casimir effect based on classical ray optics.  We show that a
    very useful approximation to the Casimir force between arbitrarily
    shaped smooth conductors can be obtained from knowledge of the
    paths of light rays that originate at points between these bodies
    and close on themselves.  Although an approximation, the optical
    method is exact for flat bodies, and is surprisingly accurate and
    versatile.  In this paper we present a self-contained derivation
    of our approximation, discuss its range of validity and
    possible improvements, and work out three examples in detail.  The
    results are in excellent agreement with recent precise numerical
    analysis for the experimentally interesting configuration of a sphere
    opposite an infinite plane.
\end{abstract}
\vspace*{-\bigskipamount} \preprint{MIT-CTP-3502} \maketitle
\section{Introduction}
\setcounter{equation}{0} Revolutionary new experimental techniques
have made possible precise measurements of Casimir
forces\cite{expt1}. Casimir's original prediction for the force
between grounded conducting plates due to modifications of the
zero point energy of the electromagnetic field has already been
verified to an accuracy of a few percent. Variations with the
conductor geometry and the effects of finite conductivity and
finite temperature will soon be measured as well. Progress has
been slower on the theoretical side.  Despite years of effort,
Casimir forces can only be calculated for the simplest geometries.
Beyond Casimir's original study of parallel plates\cite{Casimir},
we are only aware of useful calculations for a corrugated
plate\cite{Kardar} and for a sphere and a plate\cite{Gies03}.  The
former was obtained with functional integral techniques quite
special to that geometry and the latter was obtained by
computationally intensive numerical methods. Simple and
experimentally interesting geometries like two spheres, a finite
inclined plane opposite an infinite plane, and a pencil point and
a plane, remain elusive.  The Proximity Force
Approximation\cite{Derjagin} (PFA), which has been used for half a
century to estimate the dependence of Casimir forces on geometry,
was shown by Gies {\it et al.\/} \cite{Gies03} to deviate
significantly from their precise numerical result for the sphere
and plane. Thus at present neither exact results nor reliable
approximations are available for generic geometries. It was in
this context that we recently proposed a new approach to Casimir
effects based on classical optics\cite{pap1}. The basic idea is
extremely simple: first the Casimir energy is recast as a trace of
the Green's function; then the Green's function is replaced by the
sum over contributions from optical paths labelled by the number
of (specular) reflections from the conducting surfaces.  The
integral over the wave numbers of zero point fluctuations can be
performed analytically, leaving
\begin{equation}
\label{eq:cas1}
\cE_{\rm opt}= -\frac{\hbar
c}{2\pi^{2}}\sum_{r}(-1)^{r}\int_{\cD_{r}}d^{3}x
\frac{\Delta_r^{1/2}(x)}{\ell_r^3(x)}.
\end{equation}
Here $\ell_{r}(x)$ is the length of the closed geometric optics ray
beginning and ending at the point $x$ and reflecting $r$ times from
the surfaces.  $\Delta_{r}(x)$ is the enlargement factor of classical
optics\cite{Born,Kline}, also associated with the $r$-reflection path
beginning and ending at $x$.  $\cD_{r}$ is the subset of the domain,
$\cD$, between the plates in which $r$ reflections can occur.  The
factor $(-1)^{r}$ implements a Dirichlet boundary condition on the
plates; different boundary conditions require different factors.  Both
$\ell_{r}(x)$ and $\Delta_{r}(x)$ are very easy to compute either
analytically in simple cases, or numerically in general.
$\Delta_{r}(x)$, although well known in optics, may not be familiar in
the context of Casimir effects.  We will describe its properties in
some detail.

Eq.~(\ref{eq:cas1}) turns out to be a powerful tool to compute
Casimir effects for generic geometries, and to identify, interpret
and dispose of, divergences.  Eq.~(\ref{eq:cas1}) is not exact.
Instead it is an approximation which is valid when the natural
scales of diffraction are large compared to the scales that
measure the strength of the Casimir force.  In practice this will
typically be measured by the ratio of the separation between the
conductors, $a$, to their curvature, $R$.  Although approximate,
the optical approach is surprisingly accurate, as well as
physically transparent and versatile.  It generalizes naturally to
the study of Casimir thermodynamics, to the study of energy,
pressure, and momentum densities, to various boundary conditions,
to fermions, and to compact and/or curved manifolds. This is the
first in a series of papers intended to provide an introduction to
the optical approach to Casimir physics.  Here we will focus on
fundamentals: how to derive the optical approximation and how to
apply it to practical calculations of Casimir forces.  In later
papers in this series we study Casimir effects at finite
temperature, the calculation of local observables like the energy
density and pressure, and the generalization to conducting and
other boundary conditions.  Our first aim is to familiarize the
reader with the use of the optical approximation, since this
method of calculation is unfamiliar.  In Section II we present
some examples of the use of the optical approximation.  First we
review in more detail the treatment of parallel plates already
presented in Ref.~\cite{pap1}.  Although it is no great triumph to
rederive this classic result, the optical derivation illustrates
several characteristic features of the method: rapid convergence,
simple disposal of divergences and ease of computation, in
particular.  Next we present the case of a sphere and a plate.
This too was summarized in Ref.~\cite{pap1}. Here we concentrate
especially on the enlargement factor, both its interpretation and
how to compute it.  Also we illustrate the generic way that
divergences can be eliminated.  The numerical results we present
here are more accurate than those of Ref.~\cite{pap1}. Finally we
apply the optical method to the case of a finite plate suspended
above an infinite conducting plane -- the ``Casimir pendulum''. We
show how all reflections can be computed and how the optical
result differs from the proximity force approximation. In
collaboration with O.~Schroeder we are preparing a thorough study
of the hyperboloid (``pencil point'') near an infinite
plane\cite{SSJ}. In Section III we discuss the derivation of the
optical approximation from exact expressions for the Casimir
energy.  We show how a uniform approximation to the propagator
turns into a uniform approximation for the Casimir energy.  The
derivation illustrates the nature of the approximation and shows
the way toward improvements, which, in essence, amount to
including the effects of diffraction.  We present results for a
massive scalar field in $N$ dimensions in Section III. Higher spin
fields will be considered in a later paper of this series. We
discuss the general problem of divergences.  The Casimir energy is
generically divergent --- or more properly, it depends in detail
on the cutoffs that limit the conductivity of real materials at
high frequency.  However it is known that the Casimir \emph{force}
between rigid conductors is cutoff
independent\cite{Graham:2002xq}.  In the optical approximation the
cutoff dependent terms in the Casimir energy can easily be
isolated and shown to be independent of the separation between
conductors. They therefore do not contribute to forces and can be
dropped. Corrections to the optical approximation will bring in
new surface divergences.  In Section \ref{sec:connections} we
discuss the relation of the optical approximation to previous
works on ``semiclassical'' approximations to the Casimir
energy\cite{SandS}. In the last section we summarize our results,
discuss their implications, and mention extensions to other
interesting geometries.
\section{Three examples}
\setcounter{equation}{0}
In this section we present three examples of the use of the optical
approximation, eq.~(\ref{eq:cas1}).  Our aim is expressly pedagogical:
we want to demonstrate that this method can yield interesting and
accurate results without onerous calculations.
\subsection{Parallel plates}\label{parallelplates}
\begin{figure}
\begin{center}
\vspace{0.5cm}
%\BoxedEPSF{pp.eps scaled 700}
\asfigure{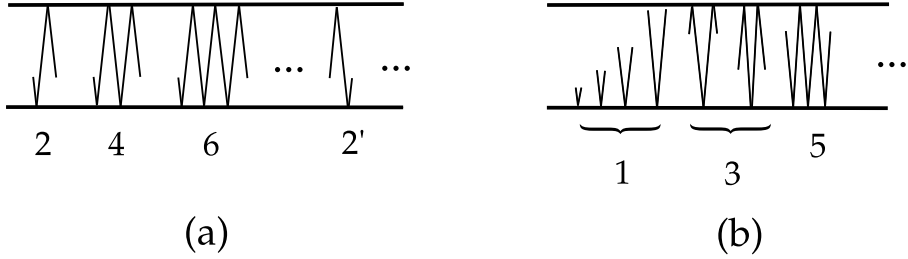}{700}{12cm}
\caption{Optical paths for parallel plates. The initial and final
points on the paths, which coincide, have been separated so the
paths can be seen.  a) Even reflections 2, 4, and 6.  Path 2' is
distinct from 2 and illustrates the origin of $M_{2n}=2$. b) Odd
reflection paths.  The paths shown form a family of continuously
increasing length.  Another family begins with the first
reflection from the top.}
\label{fig:pp}
\end{center}
\end{figure}

Casimir's original result for parallel plates can be derived in
many ways.  We present a derivation from the optical approximation
in order to illustrate several generic features of the approach in
the simplest possible context.  The points we wish to stress are:
ease of calculation; the rapid convergence in $r$, the number of
reflections; and the simple and accurate treatment of divergences.
The ``semiclassical'' method\cite{SandS} and the method of
images\cite{Brown69} generate exactly the same calculation as ours
for parallel plates. However they do not generalize to less
trivial geometries (although one might say that our method
\emph{is} the correct generalization of the method of images).
We study a massless scalar field for simplicity, and quote the
generalization to a massive scalar in a later section.  For a flat
surface the enlargement factor reduces to $1/\ell_{r}(x)$, so the
contribution of the $r$ reflection path is
\begin{equation}
\cE_r=-\frac{\hbar c}{2\pi^{2}}
(-1)^{r}M_r\int_{\cD_r}d^3x\frac{1}{\ell_r(x)^4},
\label{eq:pp}
\end{equation}
where $M_r$ is the multiplicity of the path.
It is convenient to separate the paths into ``odd'' ($r=2n+1$) and
``even'' ($r=2n$) according to the number of reflections.  Some of
these paths are shown in Fig.~[\ref{fig:pp}].  Odd and even paths
differ dramatically in their contribution to the Casimir effect:
they differ in sign and in multiplicity: $M_{r}=1$ for odd paths
and $M_{r}=2$ for even paths, as shown in the figure.  The length
of an even path depends only on $n$, whereas the length of an odd
path varies with position.  Finally, odd paths contribute a
divergence to $\cE$, but do not contribute to the Casimir force.
The even paths are finite and give the entire Casimir force.
First consider the even paths.  The length of the $2n$ reflection
path is $\ell_{2n}=2na$ independent of $x$, as can easily be seen
in Fig~[\ref{fig:pp}].  The volume of each domain, $\cD_{2n}$, is
the volume between the plates, $Sa$.  Hence the contribution from
even paths is
\begin{equation}
\cE_{{\rm even}}=-\frac{\hbar c}{2\pi^2}2Sa\sum_{n=1}^\infty
\frac{1}{(2na)^4}=-\frac{\hbar
c\pi^2}{1440a^3}S.
\end{equation}
which is the famous result due to Casimir \cite{Casimir}.
Next consider the odd paths.  There are two families.  One is
illustrated in Fig.~\ref{fig:pp}.  The other family begins with the
first reflection from the top plate.  Their contributions are
identical, giving an overall factor of two.  The $r=2n+1$ reflection
paths range in length from $2na$ to $2(n+1)a$ as can be seen from
Fig.~[\ref{fig:pp}], and contribute
\begin{equation}
    \cE_{2n+1}=\frac{\hbar
    c}{2\pi^{2}}2S\int_{na}^{(n+1)a}dz\frac{1}{(2z)^{4}},\ \mbox{for} \
    n=0,1,2,\ldots
\end{equation}
The first reflection contribution diverges at the lower limit.  As
discussed in the Introduction (and further in Section III) the
divergence indicates dependence on the properties of the material
composing the plates and is cutoff at a distance scale $\epsilon$
determined by the microphysics.  For example we can take
$\epsilon$ to be the skin depth or regard $\epsilon$ as $\sim
c/\Lambda$, where $\Lambda$ is a frequency cutoff, for example the
plasma frequency of the metal. Inserting $\epsilon$ as the lower
limit for $n=0$ and summing over $n$, we obtain the contribution
of odd paths,
\begin{equation}
    \cE_{\rm odd}=\frac{\hbar
    c}{2\pi^{2}}2S\int_{\epsilon}^{\infty}dz\frac{1}{(2z)^{4}}
    =\frac{\hbar c}{48\pi^{2}\epsilon^{3}}S
    \label{eq:ppodd}
\end{equation}
This contribution displays the cubic surface divergence expected for a
scalar field obeying a Dirichlet boundary condition\cite{Deutsch79}.
However, the divergent term --- and indeed the sum of all odd
reflections --- is independent of $a$ and therefore does not
contribute to the \emph{force} between the plates.  Until now we have
not considered the contributions from one-reflection paths that lie
below the bottom plate or above of the top plate.  It is easy to see
that the sum of these contributions is identical to
eq.~(\ref{eq:ppodd}) and does not contribute to the force.
This simple calculation illustrates some general features of the
optical approach:
\begin{itemize}
    \item The even reflections dominate, give rise to attraction,
    and their sum
    converges rapidly in $n$.  They are also attractive for Neumann
    boundary conditions, where the factor $(-1)^{r}$ is absent.  They
    would be repulsive if one surface were Neumann and the other
    Dirichlet.  In the case of parallel plates 92\% of the Casimir
    effect comes from the second reflection, 98\% from the second and
    fourth, and 99.3\% from the second, fourth and sixth reflections.
    Similar results will be found to hold in more complicated
    geometries.
    \item The only divergent contribution comes from the first reflection.
    It does not depend on the separation and therefore does not
    contribute to the Casimir force.  This result is quite general.
    To see the general argument, reconsider the first reflection from the
    bottom plate, ${\cS_{1}}$
    \begin{equation}
        \cE_{1,\cS_1}=\frac{\hbar
        c}{2\pi^2}S\int_\epsilon^a
        dz\frac{1}{(2z)^4}=\frac{\hbar
        c}{2\pi^2}S\int_\epsilon^\infty
        dz\frac{1}{(2z)^4}-\frac{\hbar
        c}{2\pi^2}S\int_a^\infty
        dz\frac{1}{(2z)^4}.
        \label{eq:self}
    \end{equation}
    The first term in eq.~(\ref{eq:self}) combined with the
    contribution of the 1-reflection path outside of the plates (from
    the lower face of the bottom  plate) is the cutoff dependent energy of
    an isolated plate.  It is manifestly independent of the presence
    of any other conductor, and gives no contribution to Casimir
    forces.  The second term is a finite effect of the first
    reflection.  For parallel plates the finite contribution of the
    first reflection is cancelled by higher odd reflections.  This
    occurs whenever the enlargement factor is $1/\ell_{n}^{2}$, that
    is, when all the conductors are planar.  For non-planar surfaces
    the first reflection gives a (relatively small) cutoff independent
    contribution to the force.
    \item The optical approach gives the exact answer for infinite
    plates.  However it will fail when $S^{1/2}\approx a$ for the same
    reason that the capacitance of two finite, parallel metallic
    plates contains corrections of order $a^2/S$\cite{Feynmanlect}: It
    is a poor approximation to consider the electric field inside two
    far separated plates ($a\gtrsim S^{1/2}$) as constant inside and
    zero outside.  Likewise, in the same limit it is a poor
    approximation to expect the Green's function for the field $\phi$
    to have contributions only from optical paths.  The corrections,
    or edge effects, can be regarded as due to diffractive rays coming
    from the edges of the plates \cite{Keller}.  We discuss
    corrections to the optical approximation in further detail in
    Section III.
    \item The difference between even and odd paths has a fundamental
    origin, as already noticed in work on the ``semiclassical''
    approximation to the Casimir energy \cite{SandS}.  The even
    paths are truly periodic, in the sense that the momentum of the
    particle, after going around the path, returns to its initial
    value.  These are therefore the paths that according to Gutzwiller
    \cite{Gutz} contribute most to the oscillations of the density of
    states.  The connection between these paths, the oscillation of
    the density of states, and the finite part of the Casimir energy
    has been noted many times\cite{Fulling} and is exact for parallel
    plates and related geometries ({\it eg.\/} flat manifolds with
    various topologies).  However, the exactness of this result is an
    accident due to the particularly simple geometry.  For example,
    there are very simple geometries in which periodic paths do not
    exist at all ({\it eg.\/} the Casimir pendulum: a finite plane
    inclined at an angle above an infinite surface).  The relation
    between the optical approach and the ``semiclassical'' approach is
    discussed further in Section III.
\end{itemize}
\subsection{The sphere and the plane}
\label{sec:sphereplane}
\begin{figure}[th]
%\centerline{\BoxedEPSF{sphere-plane.eps scaled 400}}
\centerline{\asfigure{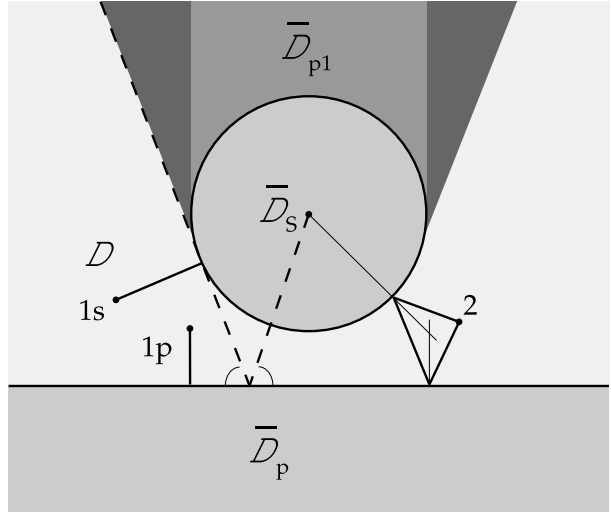}{400}{8cm}}
\caption{Geometry and reflections for a sphere and a plane. The
regions and geometrical constructions are defined in the text.}
\label{fig:sphpl}
\end{figure}

Next we analyze a problem with non-planar conductors --- typical
of real experimental configurations\cite{expt1} --- a sphere of
radius $R$ separated by a distance $a$ from an infinite plane.  In
Ref.~\cite{pap1} we tested the optical approximation by computing
the Casimir force between a sphere and a plane up through the
fourth reflection.  We showed that the optical approximation is in
very good agreement with the numerical results of
Ref.~\cite{Gies03} for $a/R \lesssim 1$.  In fact the numerical
results presented in Ref.~\cite{pap1} suffered from an
insufficiently accurate numerical integration algorithm.  The
results presented here supercede Ref.~\cite{pap1} and show that
the optical approximation is even more accurate than we originally
claimed.  For example, the optical approximation and the numerical
data differ by only 30\% at $a/R\approx 5$. Here we explain in
detail how to compute the first and second reflection
contributions.  The relevant paths are shown along with some other
aspects of the geometry in Fig.~\ref{fig:sphpl}.
For each reflection we must compute a) the optical path length,
$\ell_{r}(x)$, b) the enlargement factor, $\Delta_{r}(x)$, and c) the
domain of integration $\cD_{r}$ for which $r$-reflections are
possible.  The $\cD_{r}$ are subsets of the domain $\cD$ above the
plane and outside the sphere.
\subsubsection{Optical path lengths, $\ell_{r}(x)$}\label{paths}
Finding the $r$-reflection optical path from $x$ back to $x$,
$\ell_r(x)$, is elementary in principle.  One just draws straight
lines from $x$ to a given surface, from the arrival point on this
surface to another surface, and so on, returning after $r$ reflections
to the original point $x$.  One then moves the points of reflection on
the surfaces until one reaches the minimum total length (an elastic
string would do the job).  The minimum length path suffers specular
reflection upon each encounter with a surface.  In all but the
simplest geometries this problem must be solved numerically.  However
it is a problem amenable to extremely quick numerical solution: it is
easily defined and the minimum is unique (at least for convex
surfaces).  This procedure also defines the points of reflection,
$x_{r,1}(x)$, $x_{r,2}(x)$, etc.  The first reflection paths from the
sphere ($1s$) and the plane ($1p$) and the two reflection path are
shown in Fig.~(\ref{fig:sphpl}).
\subsubsection{Enlargement factor, $\Delta_{r}(x)$}\label{enlarge}
\begin{figure}[th]
%\centerline{\BoxedEPSF{spheregeometry.eps scaled
%650}\qquad\BoxedEPSF{enlargement.eps scaled 650}}
\centerline{\asfigure{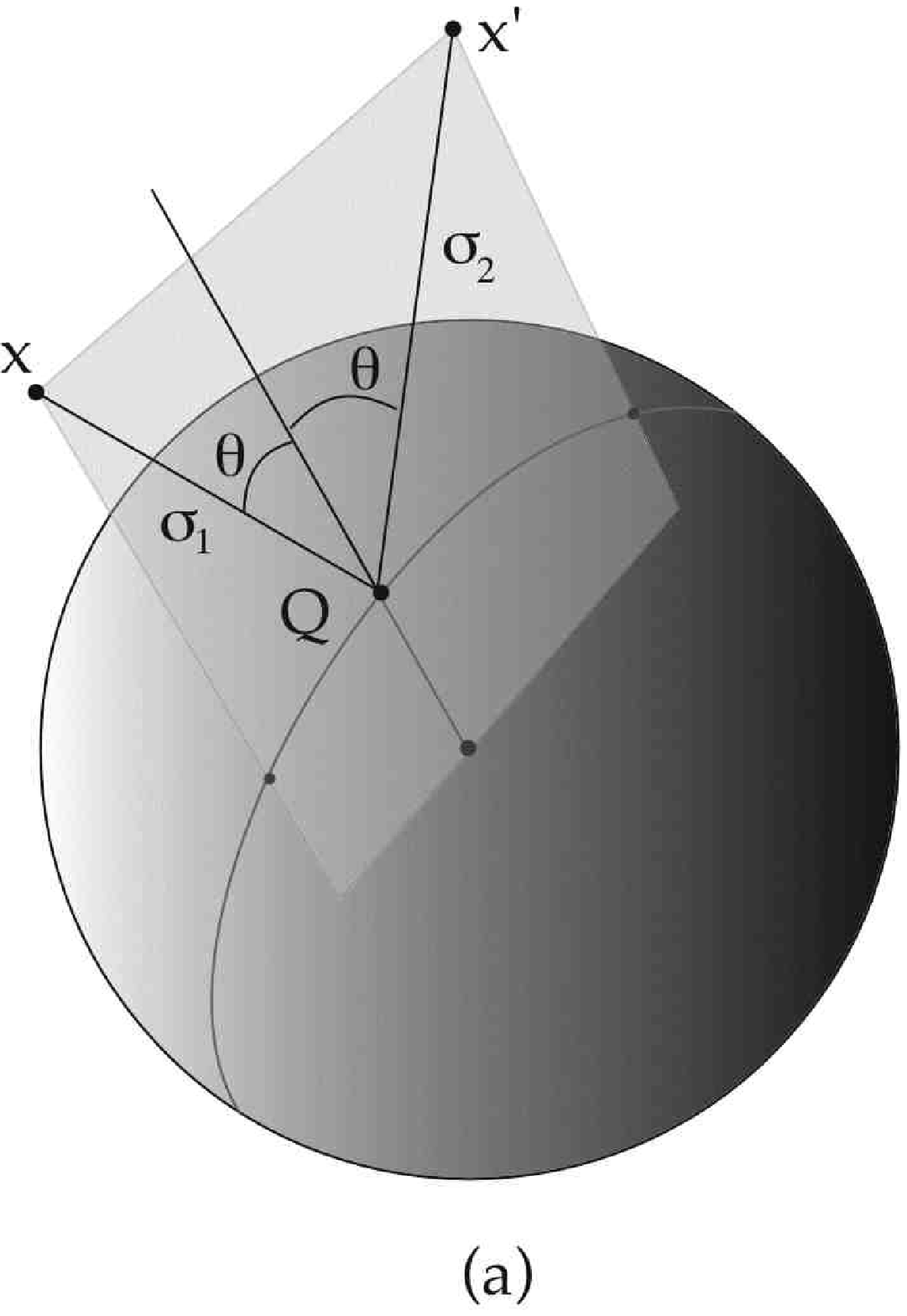}{650}{5cm}\qquad\asfigure{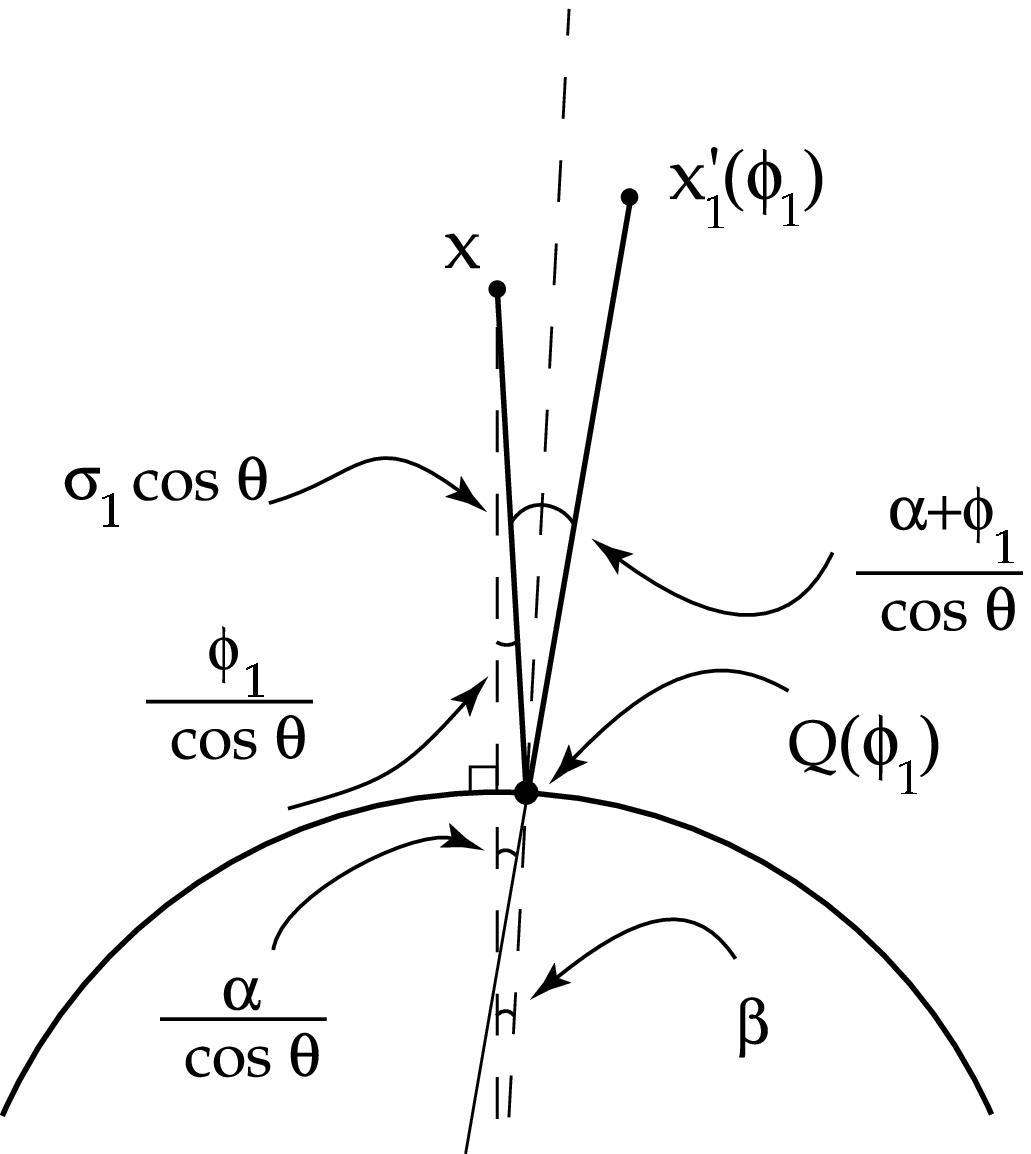}{650}{5cm}}
\caption{Geometry for reflection in a sphere. (a) The ray from $x$
to $x'$ reflecting at $Q$. Nearby rays originating at $x$ and
lying in the plane vary in longitude.  Nearby rays out of the
plane vary in latitude. (b) Variables for the calculations of the
enlargment factor associated with latitude.  The $xx'$ plane has
been projected along the vertical.  A nearby ray originating at
$x$ heading out of the $xx'Q$ plane by an angle $\phi_{1}$ is
shown. This ray reflects from the sphere at $Q(\phi_{1})$.    The
angle subtended by $x$ and $Q(\phi_{1})$ from the center of the
sphere is $\beta$. The angle formed by the vector from the center
of the sphere to $x$ and the ray from $Q(\phi)$ to
$x'_{1}(\phi_{1})$ is $\alpha$.  In the diagram the distance
$\sigma_{1}$ and the angles $\alpha$ and $\phi_{1}$ are modified
by factors of $\cos\theta$ due to the projection.}
\label{fig:enlarge}
\end{figure}

The enlargement factor for the closed path beginning and ending at
$x$ is a special case of the general enlargement factor,
$\Delta_{r}(x,x')$ for propagation from $x$ to $x'$, well known in
optics\cite{Born,Kline}.  In another guise, it is also well known
to field theorists: $\Delta_{r}(x,x')$ is just the van Vleck
determinant arising from the Gaussian fluctuations of the action
about the classical $r$-reflection path from $x$ to $x'$. In
Section III, where we discuss the origins of the optical
approximation, we show that the evaluation of the determinant
gives the standard optics definition,
\begin{equation}
\label{eq:domegada}
\Delta_r(x,x')=\frac{d\Omega_x}{dA_{x'}}
\end{equation}
From this definition it is clear that in order to obtain
$\Delta_{r}(x,x')$ one must follow the spread of an infinitesmal
pencil of rays of opening $d\Omega_{x}$ from their origin at $x$,
along this path, and measure the spread in area $dA_{x'}$ when
they arrive  at $x'$.  Having already identified the points of
reflection in  \ref{paths} it is relatively easy to compute
$\Delta(x,x')|_{x'=x}$ numerically by tracing the paths of a few
nearby rays\cite{SSJ}. It is also possible to solve this problem
analytically.  Here we present the analytic solution for the first
and second reflections from a sphere and plane.  Beyond this
level, it is probably more efficient to proceed numerically.  One
reflection from the plane is trivial:
$\Delta(x,x)=1/\ell^{2}_{1}(x)$ One reflection from the sphere is
simplified by a) normal incidence, and b) $x=x'$.  The second
reflection can be simplified by regarding it as a single
reflection from the sphere starting from $x$ and ending at the
\emph{image} $\tilde x$ of the original point $x$ in the plane. In
that case we need $\Delta_{1}(x,\tilde x)$. Consider the path from
$x$ to the sphere at the point $Q$, and then to $x'$.  To obtain
$\Delta$ one must follow the wavefront radii of curvature along
the ray. We consider a ray that impacts the sphere at an angle
$\theta$ to the normal.  It travels a distance $\sigma_{1}$ before
and $\sigma_{2}$ after, with $\ell=\sigma_{1}+\sigma_{2}$. These
variables are defined in Fig.~\ref{fig:enlarge}(a). Consider a
pencil of rays originating at $x$, spanning two infinitesmal arcs
of angular widths $d\phi_{1,2}$ along perpendicular directions.
Let $dx_{1}'$ and $dx_{2}'$ be the associated arc lengths observed
at $x'$.  Then
\begin{equation}
\Delta_r(x)=\left.\frac{d\Omega_x}{dA_{x'}}\right|_{x'=x} =\left.
\frac{d\phi_{1}}{dx_{1}'}\frac{d\phi_{2}}{dx_{2}'}\right|_{x'=x}.
\end{equation}
Since both the initial ray and the sphere have equal radii of
curvature we have the freedom to choose the directions defining
$d\phi_{1}$ and $d\phi_{2}$.  We choose ``latitude" and
``longitude" as follows:  Latitude is the
direction perpendicular to the plane formed by $x$, $x'$ and the
center of the sphere (see Fig.~\ref{fig:enlarge}(a)).  Longitude is the
direction in the plane.
Consider the pencil of rays of varying latitude as shown in
Fig.~\ref{fig:enlarge}(b).  The variables are defined in the figure.
It is easy to see that
\begin{equation}
 {dx_{1}'}= {\sigma_1d\phi_{1}+\sigma_2d\alpha},
\end{equation}
and considering that $d\alpha=2d\beta\cos\theta+d\phi_{1}$ and
$Rd\beta=\sigma_1 d\phi_{1}$, we find
\begin{equation}
\frac{d\alpha}{d\phi_{1}}=1+\frac{2\sigma_1\cos\theta}{R},
\end{equation}
and hence
\begin{equation}
\frac{d\phi_{1}}{dx_{1}'}=\frac{1}{
\ell+\frac{2\sigma_1\sigma_2\cos\theta}{R}}
\end{equation}
The same calculation applies for the longitudinal displacement
except that the relation between $d\phi_{1}$ and $d\beta$ is
replaced by $ Rd\beta = \ \sigma_1d\phi_{2}/\cos\theta$ and
$d\alpha=2d\beta+d\phi_1$, with the result
\begin{equation}
    \frac{d\phi_{2}}{dx_{2}'}=\frac{1}{\ell+
    \frac{2\sigma_1\sigma_2}{R\cos\theta}}
\end{equation}
Putting these formulas together we find for a single reflection on the
sphere (the subscript $s$ indicates reflection from the sphere) with
angle of reflection $\theta$
\begin{equation}
\Delta_{1s}(x,x')=\frac{1}{(\ell +\frac{2\sigma_1\sigma_2}{R\cos\theta})
(\ell +\frac{2\sigma_1\sigma_2\cos\theta}{R})}.
\label{deltasphere}
\end{equation}
Note that $\Delta(x,x')$ is symmetric with respect to
the interchange of $x$ and $x'$ as it must, because the
propagator possesses this symmetry.
For the first reflection from the sphere we have $\cos\theta=1$ and
$\sigma_{1}=\sigma_{2}=\ell/2$, so
\begin{equation}
    \Delta_{1s}(x)=\frac{1}{(\ell+\ell^{2}/2R)^{2}}
    \label{delta1s}
\end{equation}
and, as mentioned above, the enlargement factor for the second
reflection  (on
the sphere and then on the plane or {\it vice versa\/}), is given by
the first reflection  from $x$ to its image $\tilde x$ in the plane,
$\Delta_2(x)=\Delta_{1s}(x,\tilde{x})$.  A similar approach to higher
reflections would require further analysis.  The original wavefront
leaving $x$ is spherical.  The first reflection from the sphere
produces a new wavefront with, in general, two unequal radii of
curvature.  When next incident upon the sphere, the asymmetric
wavefront will be transformed in a manner yet to be described.  The
ease with which $\Delta_{r}(x)$ can be computed numerically makes this
unnecessary.
\subsubsection{Domain of the $r^{\rm th}$ reflection, $\cD_{r}$}
\label{domain}
The next step is the integration over the domains appropriate to each
reflection.  The first reflections give rise to cutoff dependent but
$a$-independent contributions which must be analyzed at this point.
Consider the first reflection from the plane.  The appropriate domain
is all of space except the interior of the sphere $\overline{\cD}_s$
and the region shadowed by the sphere $\overline{\cD}_{p1}$ (see
Fig.~\ref{fig:sphpl}).  The integral can then be written as the
difference between the integral over all space and the integral over
$\overline{\cD}_s\bigcup\overline{\cD}_{p1}$.  The integral over the
whole space is the divergent constant discussed in
\ref{parallelplates} which does not contribute to the force.  It is to
be ignored in the following.  So the correct domain for the first
reflection from the plane is the region,
$\overline{\cD}_s\bigcup\overline{\cD}_{p1}$, in the shadow of the
sphere, \emph{and the sign is to be reversed.}
Similarly, the integral of the first reflection on the sphere must be
performed on the domain consisting of the whole space minus the
interior of the sphere ($\overline{\cD}_{s}$) and the region below the
plate ($\overline{\cD}_{p}$).  The irrelevant divergence is given by
the integral over all the space minus the interior of the sphere and
the finite, $a$-dependent part, which contributes to the Casimir
force, is given by the negative of the integral over the region,
$\overline{\cD}_{p}$, below the plane.  So the correct domain for the
first reflection from the sphere is $\overline{\cD}_{p}$ \emph{and the
sign of the contribution is reversed.} Hence we can write
\begin{equation}
\cE_{1s}+\cE_{1p} =-\frac{\hbar c}{2\pi^{2}}
\int_{\overline{\cD}_s\bigcup\overline{\cD}_{p1}}d^3x \frac{1}{(2z)^{4}}-\frac{\hbar
c}{2\pi^{2}} \int_{\overline\cD_{p}}d^3x
\frac{\Delta_{1s}^{1/2}(x)}{\ell_{1s}^3(x)}.
\end{equation}
The second reflection gives a finite contribution to $\cE$.  The path
length, $\ell_{2}(x)$, never vanishes so there are no divergences at
short distances, and the integrand, $\Delta_{2}^{1/2}(x)/\ell^{3}(x)$,
falls rapidly at large distances.  The result is typically
approximately $90\%$ of the total result.
Higher reflections can be analyzed in a similar fashion. The
integration domains become progressively more restricted.  For
example, three reflection paths that reflect twice from the plane
and once from the sphere do not exist in the shadow of the sphere
($\overline{\cD}_{p1}$) nor in the darkly shaded regions in
Fig.~\ref{fig:sphpl} determined by the geometrical construction
indicated by the dashed lines.  It is not hard to carry out the
constructions and calculations necessary to construct the optical
approximation for the sphere and plane to any required order.
\subsubsection{Discussion of numerical results}
\begin{figure}[th]
% \centerline{\BoxedEPSF{esph.eps scaled 500}}
\centerline{\asfigure{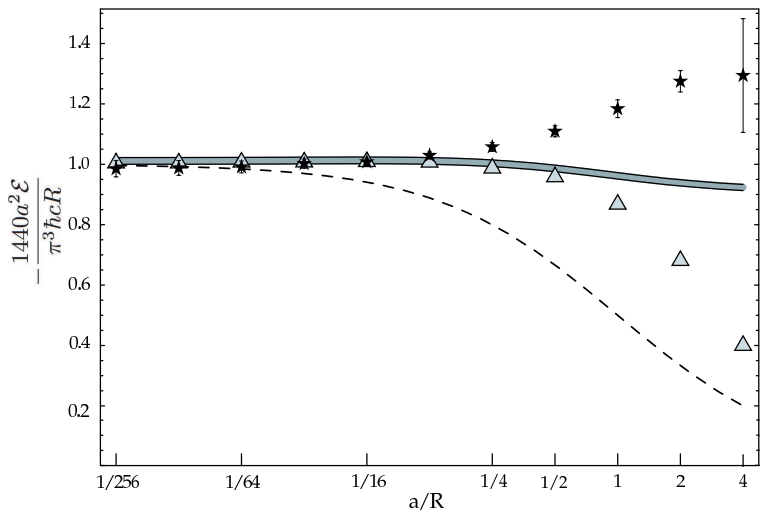}{600}{10cm}}
\caption{Casimir energy for a sphere of radius $R$ a distance $a$
above an infinite plane. $1440a^{2}\cE/\pi^{3}R\hbar c$ is plotted
versus $a/R$.  The stars with error bars are the data of
Ref.~\cite{Gies03}.  The thick solid curve is the optical
approximation through the fourth reflection.  The width of the
curve indicates our estimate of the error in the optical
approximation from neglect of the odd and even reflections with
$n\ge 5$.  The dashed curve is the plate-based proximity force
approximation.  The triangles are the results we published in
Ref.~\cite{pap1}, which are superceded by this work.}
\label{fig:esph}
\end{figure}

\begin{figure}[th]
%\centerline{\BoxedEPSF{pieces.eps scaled 600}}
\centerline{\asfigure{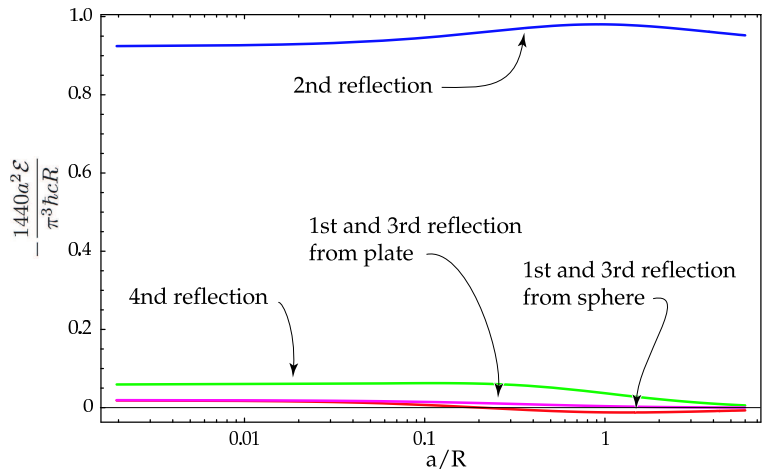}{600}{10cm}}
\caption{Contributions of specific reflections to the optical
approximation. }
\label{fig:pieces}
\end{figure}

In Ref.~\cite{pap1} we presented initial results on the optical
approximation for the sphere and plane.  Here we present final
results (see Fig.~\ref{fig:esph}), discuss them in more detail,
and compare them with the results of Ref.~\cite{Gies03} and with
the proximity force approximation (PFA).  In presenting our
results we display the sum of all the reflections (even and odd)
up to (and including) the fourth. Since the energy must approach
the parallel plate limit as $a\to 0$ we can estimate the error in
neglecting higher reflections in this limit.  The error in
neglecting the fifth and higher odd reflections is a $+3.8\%$
excess (because the sign of the odd reflections contribution is
opposite to that of the total energy) as $a\to 0$.  Neglecting the
even reflections (6th, 8th, etc.)  as $a\to 0$ gives an error of
$-1.8\%$, negative because these contributions have the same sign
of the total energy.  Altogether the sum of the first four
reflections overestimates the energy by $3.8\%-1.8\%=2\%$ as $a\to
0$.  To illustrate this estimate of accuracy we have plotted our
results as a band $2\%$ in width in Fig.~\ref{fig:esph}.  Since
the fractional contribution of higher reflections decreases with
$a$, we believe this is a conservative estimate for larger $a$.
Obviously, calculating the higher reflections will reduce this
uncertainty interval, leaving eventually only the error due to
diffraction which we are not able to estimate. The proximity force
approximation has been the standard tool for estimating the
effects of departure from planar geometry for Casimir effects for
many years\cite{MT}.  In this approach one views the sphere and
plate as a superposition of infinitesimal parallel plates:
\begin{equation}
    \cE_{\rm PFA} = -\frac{\pi^{2}\hbar c}{1440}\int_{\cS}
    d^{2}S\frac{1}{d(x)^{3}}.
    \label{PFA}
\end{equation}
where $d(x)$ is the distance from
the plate to the sphere at a point $x$ on the surface $\cS$.
This formulation is ambiguous since the surface $\cS$ could be taken
to be either the sphere or the plate.  Whichever surface is chosen,
the distance is measured normal to that surface.  The ambiguity is
useful since it gives a measure of the uncertainty in the PFA.  In
either case the relevant integrals are easily performed.  For the plate
we obtain,
\begin{equation}
    \cE_{\rm PFA}^{\rm plate} = -\frac{\pi^{3}\hbar cR}{1440 a^{2}} \frac{1}{1+a/R}
    \label{PFApl}
\end{equation}
while for the sphere we obtain
\begin{equation}
    \cE_{\rm PFA}^{\rm sphere} = -\frac{\pi^{3}\hbar cR}{1440 a^{2}}
    \left(1-3\frac{a}{R}-6\frac{a^{2}}{R^{2}}\left(1+(1+\frac{a}{R})
    \ln\frac{a}{R+a}\right)\right)
    \label{PFAsph2}
\end{equation}
In the limit $a/R\to 0$ both estimates agree to lowest order. The
$a\to 0$ limit is usually called the proximity force
\emph{theorem} and has been much discussed over the years.  It is
usually stated as a result for the Casimir \emph{force} in the
limit $a/R\to 0$: $ \cF_{\rm PFA}\sim 2\pi R \cE/A = -
\pi^{3}\hbar c R/720 a^{3}$ (where $\cE/A$ is the Casimir energy
per unit area for parallel plates).  This limit provides a
convenient parametrization of the Casimir force when $a$ is not so
small,
\begin{equation}
    \cF  = -f(\frac{a}{R})\frac{\pi^{3}\hbar c R}{720a^{3}}
    \label{ffunction}
\end{equation}
Modern experiments are approaching accuracies where the deviations
of $f(a/R)$ from unity may be important.  The accuracy of the PFA
beyond the $a/R\to 0$ limit is unknown, and the two different versions
give different $\cO(a/R)$ corrections:
\begin{eqnarray}
    f_{\rm PFA}^{\rm plate}(a/R) &=& 1-\frac{1}{2}\frac{a}{R} + {\cal
    O}(\frac{a^{2}}{R^{2}})
\label{PFAforcePlate}
\\    f_{\rm PFA}^{\rm sphere}(a/R) &=& 1-\frac{3}{2}\frac{a}{R} +
{\cal
    O}(\frac{a^{2}}{R^{2}})
    \label{PFAforceSphere}
\end{eqnarray}
An important application of the optical approximation is to obtain a
more accurate estimate of $f(a/R)$.
The optical approximation to the Casimir energy and the data of
Ref.~\cite{Gies03} both fall like $1/a^{2}$ at large $a$.  In fact
both are roughly proportional to $1/a^{2}$ for all $a$.  In
contrast the PFA estimates of the energy falls like $1/a^{3}$ at
large $a$ and depart from the Gies et al data at relatively small
$a/R$.  For purposes of display we therefore scale the estimates
of the energy by the factor $-1440a^{2}/\pi^{3}R\hbar c$.  The
results are shown in Fig.~\ref{fig:esph}.
At large $a/R$ the optical approximation has the same scaling
behavior as the data and differs from Ref.~\cite{Gies03} by no
more than 30\% at the largest $a/R$.  At small $a/R$, given our
estimate of the accuracy of the optical approximation, we find
that
\begin{equation}
f^{\rm optical}(a/R)=1+0.05 a/R+\Ord{(a/R)^2},
\end{equation}
which must be compared with the predictions of PFA
eqns.(\ref{PFAforcePlate}) and (\ref{PFAforceSphere}).
In Fig.~\ref{fig:pieces} we show the contributions to the optical
approximation of the different reflections we have computed.  As
expected the dominant contribution, always greater than 90\%,
comes from the second reflection.  The fourth reflection
contributes about 6\% for $a/R\ll 1$ and less as $a/R$ increases.
The contributions of the first and third reflections are very
small for all $a/R$.  A relevant result, confirmed by the
analytical analysis on the energy momentum tensor (within the
optical approximation), the subject of the second paper of this
series, is that the asymptotic behavior of $\cE$ as $a/R\gg 1$
predicted by the optical approximation is $\propto 1/a^2$.  This
is in contrast with the Casimir-Polder law which predicts $1/a^4$
at large $a$ the discrepancy is to be attributed to diffraction
effects.
\subsection{Casimir Pendulum}
\begin{figure}[th]
%\centerline{\BoxedEPSF{pendulum.eps scaled 550}}
\centerline{\asfigure{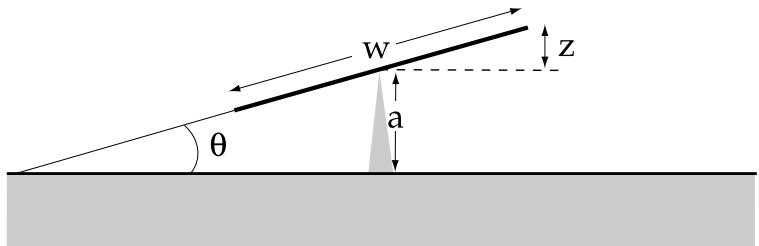}{550}{10cm}}
\caption{Geometry for the Casimir Pendulum.}
\label{fig:pendulum}
\end{figure}

\begin{figure}[th]
%\centerline{\BoxedEPSF{nooddrefl.eps scaled 550}}
\centerline{\asfigure{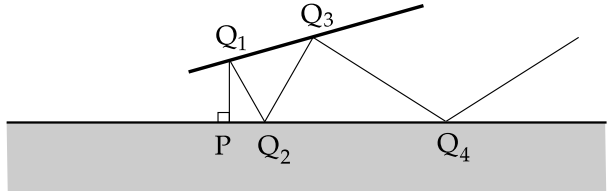}{550}{8cm}}
\caption{Odd reflection paths for the Casimir Pendulum.}
\label{nooddrefl}
\end{figure}

In this section we treat a problem for which the exact answer is
unknown.  The configuration is shown in Fig.~\ref{fig:pendulum}.
The base plate is taken to be infinite.  The upper plate is held
at its midpoint a distance $a$ above the base plate.  The width of
the upper plate is $w$ and its depth, $d$ (out of the page), is
assumed to be infinite.  We define the Casimir energy per unit
depth, $\varepsilon=\cE/d$.  $\theta$ is the angle of inclination
of the upper plate.  It will be convenient to use
$z=\tiny{\frac{1}{2}}w\sin\theta$ as a variable as well.  It is
also possible to view this configuration as a finite slice between
$\ell_{1}=a/\sin\theta -w/2$ and $\ell_{2}=a/\sin\theta+w/2$ of a
wedge of opening angle $\theta$.  In this section we will discuss
both the Casimir energy and the ``Casimir torque'',
$\nu=\frac{1}{d}\frac{d\cE}{d\theta}$, per unit depth.
We are aware of two {\it ad hoc\/} approximate approaches to this
problem.  The first is the PFA which treats each element of the system
perpendicular to the lower plate as an infinitesmal parallel plate
Casimir system.  It is easy to show that
\begin{eqnarray}
\varepsilon_{\rm PFA} &= &- \frac{\pi^{2}\hbar
c}{1440}\frac{w\cos\theta}{a^{3}}
\left(1-\frac{w^{2}\sin^{2}\theta}{4a^{2}}\right)^{-2}\nonumber\\
&=& -\frac{\pi^{2}\hbar c
a}{1440}\frac{\sqrt{w^{2}-4z^{2}}}{(a^{2}-z^{2})^{2}}
\label{pendpfa}
\end{eqnarray}
which gives a torque,
\begin{equation}
\nu_{\rm PFA}(a,w,z) = - \frac{\pi^{2}\hbar c}{720}\ \frac{az(w^{2}-a^{2}-3z^{2}
)}{(a^{2}-z^{2})^{3}}
\label{pendpfatorque}
\end{equation}
where the minus sign denotes that the torque is destabilizing:
$z=0$ is a point of unstable equilibrium.  As in the case of the
sphere and the plane, the PFA is ambiguous.  A more symmetric
treatment of the two planes in the present geometry would
integrate over the surface that bisects the wedge and take the
distance normal to that surface. The result is the replacement of
$\cos\theta$ by $\cos^{4}(\theta/2)$ in eq.~(\ref{pendpfa}) and a
similar modification of the torque.
A second ``approximate'' treatment of the Casimir Pendulum can be
extracted from the known exact solution for the Casimir energy
density for the ``Dirichlet wedge''\cite{dirichletwedge}, which
consists of two semi-infinite plates with opening angle $\theta$
meeting at the origin.  One can obtain an estimate of the energy
for the pendulum by integrating the energy density over the two
dimensional domain bounded (in polar coordinates, $(\rho,\phi)$)
by $0<\phi<\theta$ and $\ell_{1}<\rho<\ell_{2}$.  This approach
takes no account of the modification of the energy density due to
the finitenss of the upper plate.  Furthermore it is inherently
ambiguous because the energy density for a scalar field is only
defined up to a total derivative. The calculation in
Ref.~\cite{dirichletwedge} used the conformally invariant stress
tensor.  One would obtain a different answer if one used, for
example, the Noether stress tensor.  In light of these
difficulties, we do not pursue this approach further. To compute
the optical approximation we need the enlargement factor, the
lengths of optical paths, and the integration domain, $\cD_{r}$.
Since all the conducting surfaces are planar, the enlargement
factor is trivial in this case, $\Delta_{r}(x)\to
1/\ell_{r}^{2}(x)$.  The path lengths are also easy to compute.
The only non-trivial step is the determination of the integration
domains. As in the case of parallel plates, the odd reflections do
not contribute to forces or torques for the Casimir pendulum.
Instead they sum to a cutoff dependent constant associated with
each plate. Any odd reflection path ``turns around'' with a
reflection at normal incidence from one plate or the other.
Consider all the points, $x$, which are the origins of paths that
turn around at a given point $P$ on either surface.  These paths
are shown, for the case where $P$ is on the lower, infinite plane,
in Fig.~\ref{nooddrefl}.  They comprise one reflection paths lying
on the interval $\overline{PQ_{1}}$, three reflection paths lying
on the interval $\overline{Q_{1}Q_{2}}$, {\it etc.\/} The
contributions to $\varepsilon$ from these intervals integrates to
the same result as the integral over $z$ for odd paths in the case
of parallel plates.  It is independent of $a$, $w$, and $\theta$
and can be set aside.  The fact that the enlargement factor is
trivial is crucial for this argument.
\subsubsection{Even optical path lengths, $\ell_{r}(x)$}\label{pendulumpaths}
\begin{figure}[th]
% \centerline{\BoxedEPSF{evenrefl.eps scaled 550}}
\centerline{\asfigure{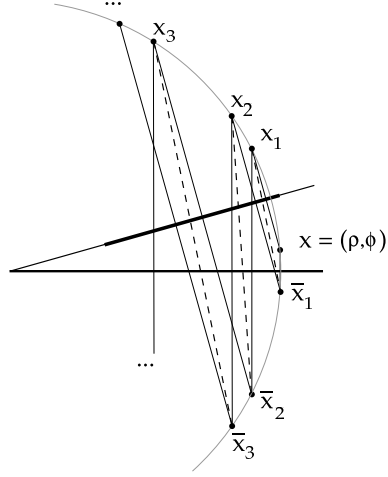}{550}{5cm}}
\caption{Images of the point $x$ in the Casimir Pendulum
configuration.  The dashed lines have the same length as the
$r=2,4,6,\ldots$ reflection paths.}
\label{fig:evenrefl}
\end{figure}

The analysis of paths that reflect an even number of times makes use
of simple geometrical concepts.  For any point $x\equiv(\rho,\phi)$,
we define an infinite sequence of images in the upper and lower planes
as shown in Fig.~\ref{fig:evenrefl}, ignoring for the moment that the
upper plane is finite.  The images below the lower plane are denoted
$\bar x_{1},\bar x_{2},\ldots$ and those above the upper plane are
denoted $x_{1},x_{2},\ldots$.  In sequence, $\bar
x_{1}={\mathbf{\overline R}}x$, $x_{1}=\mathbf{R}x$,
$x_{2}=\mathbf{R}\bar x_{1}$, $\bar x_{2}=\mathbf{\overline R} x_{1}$,
{\it etc.\/} with $\mathbf{\overline R}$ denoting reflection in the
lower plane and $\mathbf{R}$ denoting reflection in the upper plane.
All of the images lie on the circle of radius $\rho$ about the origin.
The length of the $2n$ reflection path can easily be seen to be given
by
\begin{equation}
 \ell_{2n}(x)=||x_{n}-\bar x_{n}||=4\rho^{2}\sin^{2} n\theta
 \label{pendulumlength}
\end{equation}
independent of $\phi$.
Substituting into eq.~(\ref{eq:cas1}) we obtain the expression for the
Casimir energy per unit depth,
\begin{equation}
    \varepsilon_{\rm opt} = -\frac{\hbar c}{16\pi^{2}}\sum_{n=1}^{N_{\rm
    max}}\frac{1}{\sin^{4}n\theta} \int_{0}^{\theta}
    d\phi\int_0^\infty
    d\rho\frac{1}{\rho^{3}}\  \Theta (\cD_{2n})
    \label{pendulum1}
\end{equation}
The step function $\Theta(\cD_{2n})$ vanishes when the point $x$ is not
in the domain where $2n$-reflection paths are possible.  As we show in
the following section, for $n$ large enough, $\cD_{2n}\to \emptyset$.
$N_{\rm max}$, the upper limit on the $n$-sum, is the largest
value of $n$ for which any $2n$-reflection paths exist.
\subsubsection{Domain of the $2n^{\rm th}$ reflection, $\cD_{2n}$}
The domain in which the $2n$-reflection path exists is determined
by the constraint that the points of reflection at the upper plate
must lie between $\ell_{1}$ and $\ell_{2}$, the inner and outer
radii that define its boundaries.  Note, of course, that
$\ell_{2}>\ell_{1}$. Although the calculation is elementary, it is
tricky, so we only quote the results.  The constraints depend on
whether $n$ is even or odd, so we summarize them independently.
\begin{itemize}
    \item {\bf $n$-odd}
    When $n$ is odd, the integration domain $\cD_{2n}$ is defined by the
    inequalities:
    \begin{equation}
        \ell_{1}\cos\phi \le \rho\cos n\theta \le \ell_{2}\cos
        ((n-1)\theta+\phi)
        \label{rhooddlimit}
    \end{equation}
    where the lower limit ensures that the innermost reflection occurs
    at $\rho\ge\ell_{1}$ and the upper limit ensures that the outermost
    reflection occurs at $\rho\le\ell_{2}$.  The inequality cannot be
    satisifed  for any $\rho$ or $\phi$($\le\theta$) unless
    \begin{equation}
        \ell_{2}\cos(n-1)\theta > \ell_{1}
        \label{mainodd}
    \end{equation}
    When eq.~(\ref{mainodd}) is satisfied
    the contribution of the
    $2n^{\rm th}$ reflection ($n$-odd) is
    \begin{eqnarray}
        \varepsilon^{\rm odd}_{{\rm opt}\ n} &=&-\frac{\hbar c}{32\pi^{2}}
        \frac{\cos^{2}n\theta}{\sin^{4}n\theta}\left\{\matrix{
        \frac{1}{\ell_{1}^{2}}\tan\theta
        -\frac{1}{\ell_{2}^{2}}\left(\tan n\theta -\tan(n-1)\theta\right)
        & \mbox{for}\ \  \ell_{2}\cos n\theta \ge \ell_{1}\cos\theta
        \cr
        \frac{(\ell_{2}\cos(n-1)\theta-\ell_{1})^{2}}
        {\ell_{1}^{2}\ell_{2}^{2}\ \cos(n-1)\theta\ \sin (n-1)\theta}
        \qquad \ \ \qquad&
        \mbox{for}\ \ \ \ell_{2}\cos n\theta \le \ell_{1}\cos\theta}\right.
    \label{nodd}
    \end{eqnarray}
        \item {\bf $n$-even}
    When $n$ is even, the integration domain $\cD_{2n}$ is defined by the
    inequalities:
    \begin{equation}
        \ell_{1}\cos(\theta-\phi) \le \rho\cos n\theta \le \ell_{2}\cos
        ((n-1)\theta+\phi)
        \label{rhoevenlimit}
    \end{equation}
    where, as before, the lower limit ensures that the innermost
    reflection occurs at $\rho\ge\ell_{1}$ and the upper limit ensures
    that the outermost reflection occurs at $\rho\le\ell_{2}$.  The
    inequality cannot be satisifed at all unless
    \begin{equation}
        \ell_{2}\cos(n-1)\theta > \ell_{1}\cos\theta
        \label{maineven}
    \end{equation}
    When eq.~(\ref{maineven}) is satisfied the contribution of the
    $2n^{\rm th}$ reflection ($n$-even) is
    \begin{eqnarray}
        \varepsilon^{\rm even}_{{\rm opt}\ n}&=&-\frac{\hbar c}{32\pi^{2}}
        \frac{\cos^{2}n\theta}{\sin^{4}n\theta}\left\{\matrix{
        \frac{1}{\ell_{1}^{2}}\tan\theta
        -\frac{1}{\ell_{2}^{2}}\left(\tan n\theta -\tan(n-1)\theta\right)
        & \mbox{for}\ \  \ell_{2}\cos n\theta \ge \ell_{1}
        \cr
        \frac{(\ell_{2}\cos(n-1)\theta-\ell_{1}\cos\theta)^{2}}
        {\ell_{1}^{2}\ell_{2}^{2}\ \cos(n-1)\theta\ \sin n\theta\ \cos \theta}
        \qquad \ \ \qquad&
        \mbox{for}\ \  \ell_{2}\cos n\theta \le \ell_{1} }\right.
    \label{neven}
    \end{eqnarray}
\end{itemize}
The torque is obtained by differentiating with respect to $\theta$
at fixed $a$ and $w$, remembering that $\ell_{1}$ and $\ell_{2}$
depend on $\theta$.   Of course the $\theta$-derivative of
eqs.~(\ref{neven}) and (\ref{nodd}) are complicated and need not
be written down explicitly.  The resulting expressions for
$\nu^{\rm odd}_{{\rm opt}\ n}$ and $\nu^{\rm even}_{{\rm opt}\ n}$
must be summed over $n$ subject to the constraints in
eqs.~(\ref{nodd}) and (\ref{neven}).  This sum must be performed
numerically.  The results are discussed in the following
subsection.
\subsubsection{Discussion}
\begin{figure}[th]
%\centerline{\BoxedEPSF{pendenergy.eps scaled 550}\qquad
%\BoxedEPSF{pendtorque.eps scaled 580}}
\centerline{\asfigure{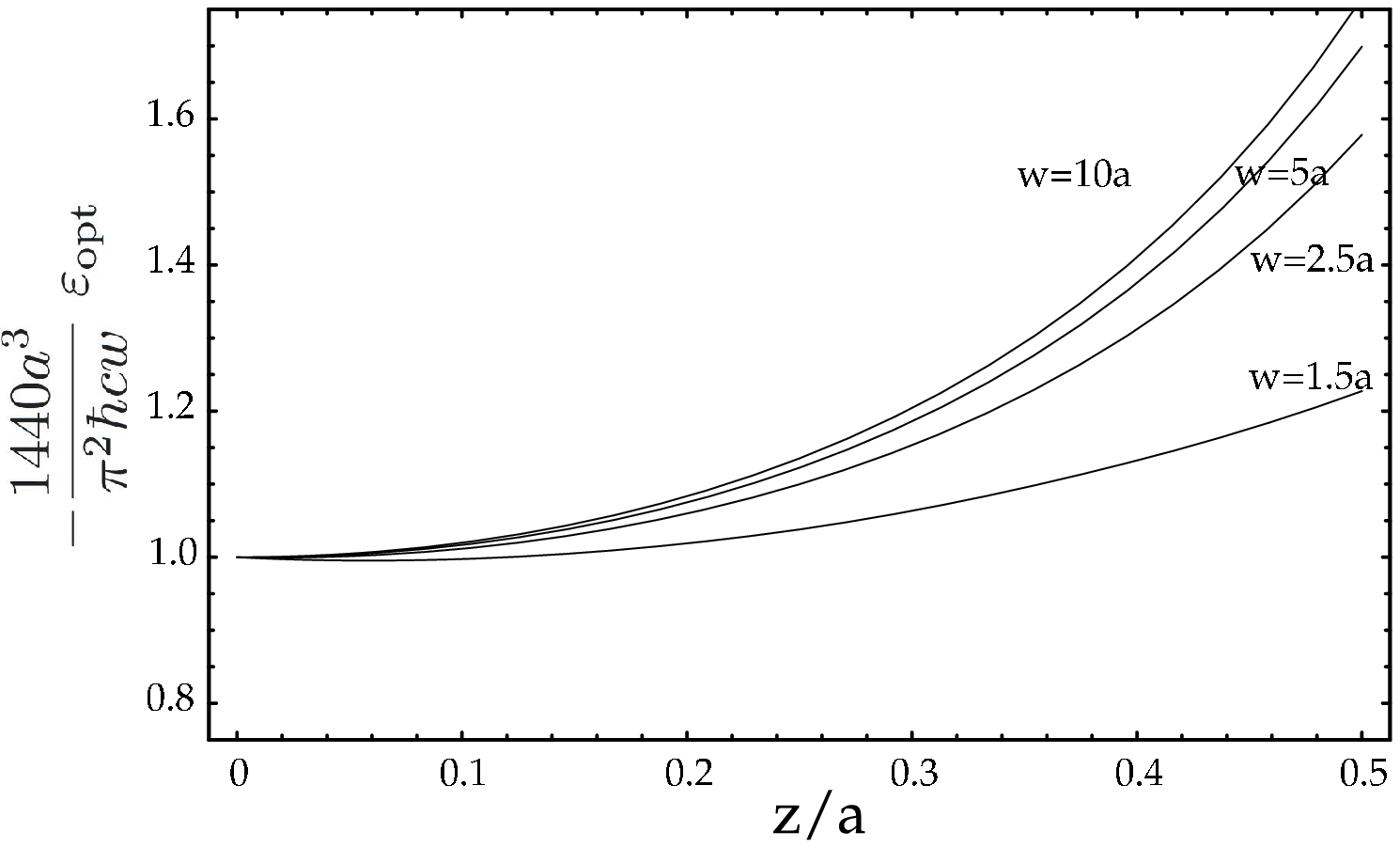}{550}{7cm}\qquad\asfigure{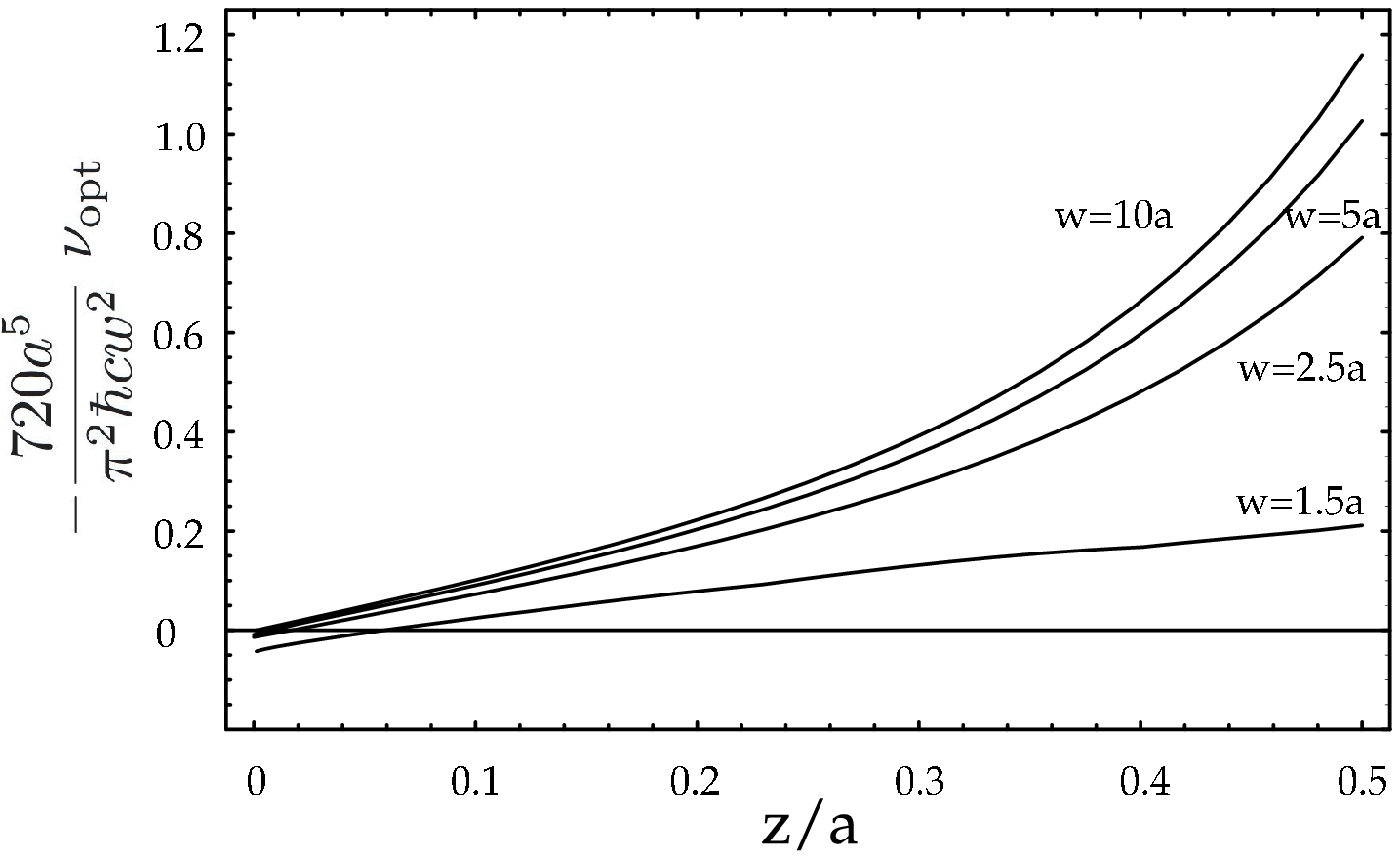}{580}{7cm}}
\caption{Casimir energy and torque in scaled units for a Casimir
pendulum of width $w=1.5a,2.5a, 5a$ and $10a$.  Positive values of
the scaled torque are destabilizing.}
\label{fig:pendenergy}
\end{figure}

\begin{figure}[th]
%\centerline{
%\BoxedEPSF{ratio.eps scaled 578}\qquad\BoxedEPSF{piecesofpendulum.eps scaled
%580}}
\centerline{\asfigure{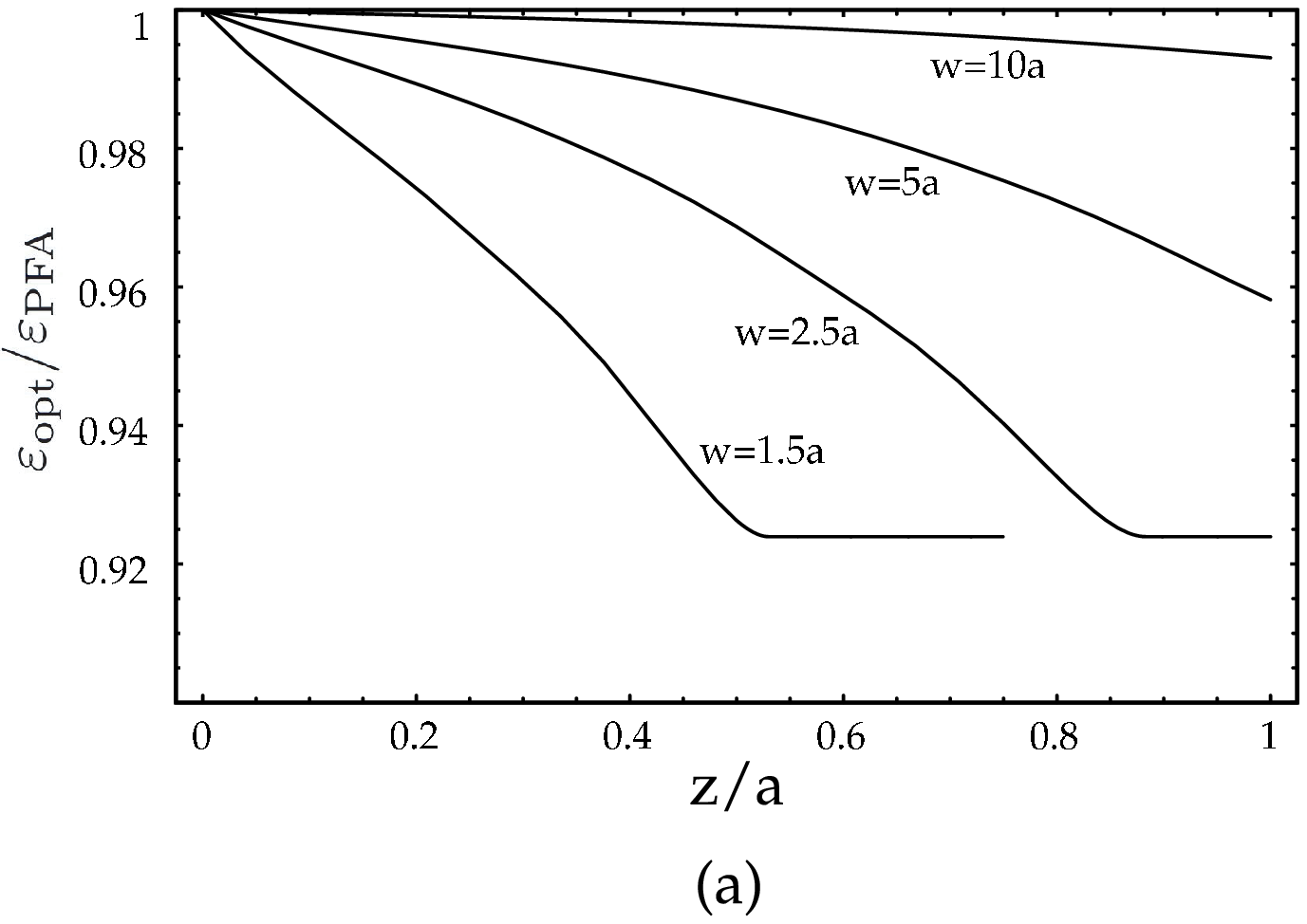}{578}{7cm}\qquad\asfigure{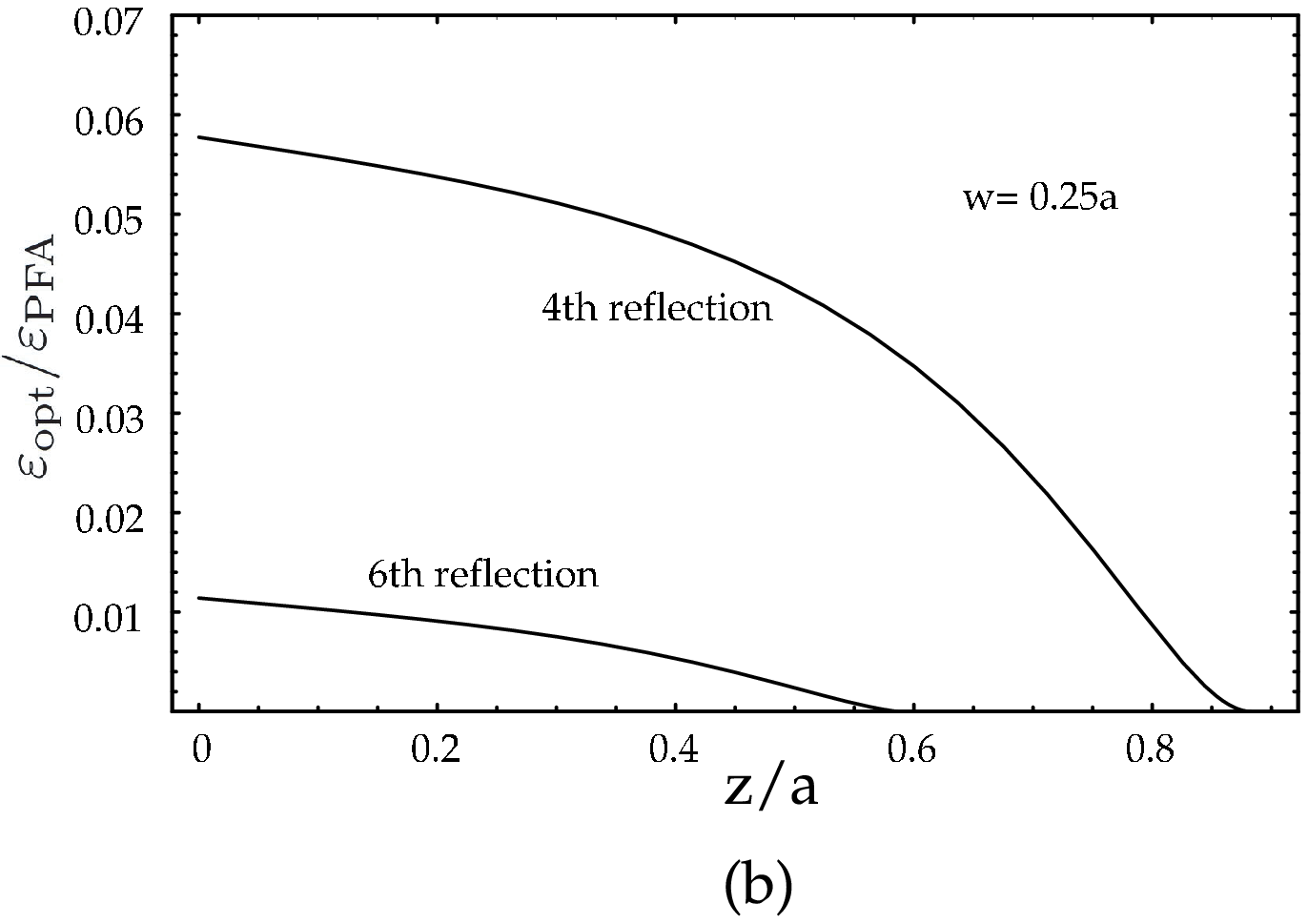}{580}{7cm}}
\caption{The ratio of the  optical approximation to the PFA. (a)
for a pendulum of different widths as a function of $z/a$. The
breaks in the curves for $w=1.5a$ and $2.5a$ occur when only the
second reflection can contribute.  The $w=1.5a$ curve ends at the
$z=0.75a$ when $z=w/2$. (b) The contributions of the 4th and 6th
reflections for $w=2.5a$.  The 2nd reflection contributes
0.924\ldots independent of $z$.  This is the only case we found in
which the optical approximation gives an energy smaller than the
PFA.}
\label{fig:ratiowedge}
\end{figure}

\begin{figure}[th]
%\centerline{ \BoxedEPSF{negative.eps scaled 600} }
\centerline{\asfigure{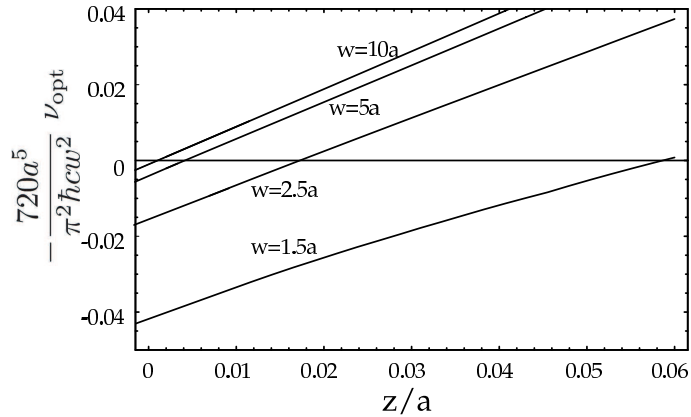}{600}{9cm}}
\caption{Torque in scaled units for a Casimir pendulum of width
$w=1.5a,2.5a, 5a$ and $10a$, for $z\approx 0$. Positive values of
the scaled torque are destabilizing.}
\label{fig:negative}
\end{figure}

Fig.~\ref{fig:pendenergy} shows the pendulum Casimir energy as a
function of $z$ for $a=1$ and several values of $w$.  The weak
dependence on $z$ at small $z$ is to be expected.  So is the
divergence as $z\to 1$ which we do not show in the figure. When
the plates touch at $z=1$, the Casimir energy of perfectly sharp,
perfectly conducting plates would in fact diverge, as would the
Casimir torque.
The optical approximation for the pendulum turns out to be very close
to the plate based PFA.  It is convenient therefore to scale out a
factor of $-1440a^{3}/\hbar c\pi^{2}w$ (see eq.~(\ref{pendpfa})) when displaying
or results for the energy per unit depth, $\varepsilon_{\rm opt}$, and
a factor $-720 a^{5}/\hbar c \pi^{2}w^{2}$ when displaying results for
the torque per unit depth.
The energy and torque are plotted for representative values of $w$
($w=1.5a,2.5a,5a,$ and $10a$) as a function of $z/a$ in
Fig.~\ref{fig:pendenergy}.  The plots are shown only up to $z/a=0.5$.
Above $z/a\sim 0.5$ both grow rapidly and diverge at $z=a$.  The PFA
gives an excellent approximation for the pendulum over a wide range of
the parameter space.  This can be seen by examining the ratio of
$\varepsilon_{\rm opt}/\varepsilon_{\rm PFA}$ as shown in
Fig.~\ref{fig:ratiowedge}.  The reason behind this success is that the
second (optical) reflection is proportional to the PFA result
(eq.~(\ref{pendpfa})) for all $z$.  The constant of proportionality is
the familiar $90/\pi^{4}=0.924$.
The sum of the higher even reflections combines with the second to
equal the PFA at $z=0$ and drops away slowly with increasing
$z/a$.  So the optical estimate coincides with the PFA at $z=0$
and drops slowly with $z$.  The contributions of the first few
reflections are shown for $w=2.5a$ in Fig.~\ref{fig:ratiowedge}.
A careful study of Fig.~\ref{fig:pendenergy} reveals one peculiar
feature of the optical approximation which is interesting in
principle, if not in practice.  The torque does not vanish at
$z=0$, and therefore the Casimir energy has a cusp at $z=0$.  The
section of the graph near $z=0$ is enlarged in
Fig.~\ref{fig:negative} to make the effect clearer.  The torque
starts negative (stabilizing) at very small $z$ before it changes
sign and becomes destabilizing as $z$ increases.  This is an
effect of the finite extent of the upper plate. It vanishes like
$a/w$ at large $w$.  We are not able to determine whether this is
an artifact of the optical approximation's neglect of diffraction,
or whether it is a real effect at small $z$.  We know of no
theorem that precludes a cusp in the energy at $z=0$.  We will
have to wait until the effects of diffraction can be included
before analyzing this further.
\section{Origins of the optical approximation}
\setcounter{equation}{0}
\label{sec:semiclapprox}
Most studies of Casimir energies do not consider approximations.
Instead they focus on ways to regulate and compute the sum over modes,
$\sum \frac{1}{2}\hbar\omega$\cite{MT}.  These methods have proved
very difficult to apply to geometries other than parallel plates.  The
main reason for this \emph{impasse} lies in the requirement of an
analytic knowledge of the spectrum.  Finding the spectrum of the
Laplace operator for non-separable problems is not merely a technical
difficulty, it is more one of principle.  In fact there are strict
relations between this problem and those of chaotic billiards theory.
The existence of an exact solution for the Casimir problem with a
non-trivial geometry would imply the existence of an exact solution
for the same family of quantum billiards and hence of classical
billiards.  30 years of work on the ergodicity of classical billiards
and the implications for the density of states in the corresponding
quantum billiards suggest this task is hopeless (see \cite{Gutz}).  On
the other hand, one could argue that a numerical solution of the
Dirichlet problem could easily give the spectrum to some high, but
finite, accuracy, and this could be used to compute the Casimir
energy.  However the Casimir energy diverges -- the leading
non-trivial divergence in $N$ dimensions is of order $\Lambda^{N}$.
The Casimir \emph{force} between rigid bodies is known to be finite,
and one could hope to compute it by introducing a cutoff, computing
the energy at nearby separations, $a$ and $a+da$, taking the
difference, $\cE(a+da,\Lambda)-\cE(a,\Lambda)$, and finally taking
$\Lambda\to\infty$.  However such numerical problems are hopelessly
unstable.  The force, indeed, is given by the small oscillatory ripple
in the density of state numerically shadowed by the `bulk'
contributions which give rise to distance-independent divergencies.
So we focused our attention on ways to get approximate solutions of
the Laplace-Dirichlet problem which are apt to capture the oscillatory
contributions in the density of states, providing physical insights
and accurate numerical estimates.  We have not found any previous use
of ideas from classical optics.  In this section we give a derivation
of the optical approximation based on a path integral representation
of the Helmholtz Greens function.  Schaden and Spruch have developed
an approximation for Casimir energies \cite{SandS} using Gutzwiller's
semiclassical treatment\cite{Gutz} of the density of states.  It is
misleading to call the approach of Ref.~\cite{SandS} ``semiclassical''
because, as can be seen for example from eq.~(\ref{eq:casimiren}), the
only $\hbar$ in the Casimir problem for a massless field is the
multiplicative factor in $\frac{1}{2}\hbar\omega$.  However, since the
authors of Ref.~\cite{SandS} use the term following Gutzwiller, we
will continue to refer to their approach as ``semiclassical''.  This
work differs in important ways from ours and in general is not as
accurate, however the relationship between the two approaches is
interesting, and is explored later in this section.
\subsection{Derivation}
We begin with the well-known definition of the Casimir energy in
terms of a space and wavenumber dependent density of
states\cite{densityofstates}, $\tilde\rho(x,k)$,
\begin{eqnarray}
    \label{eq:casimiren}
    \cE_{\cD}[\psi]=\int_0^\infty dk \int_{\cD} d^{N} x
    \frac{1}{2}\hbar\ \omega(k)\tilde\rho(x,k),
\end{eqnarray}
where $\omega(k)=c\sqrt{k^{2}+\mu^{2}}$, and the density of states
$\tilde\rho(x,k)$ is related to the propagator $G(x',x,k)$ by
\begin{equation}
    \label{eq:rhoG}
    \tilde\rho(x,k)=\frac{2k}{\pi}\ \Im \ \tilde G(x,x,k).
\end{equation}
Since we are considering a scalar field, $G$ is the Greens
function for the Helmholtz equation.  We choose $G$ to be analytic
in the upper-half $k^{2}$-plane (or equivalently take $k^{2}$ to
have a small positive imaginary part). The tildes on
$\tilde\rho(x,k)$ and $\tilde G(x,x',k)$ denote the subtraction of
the contribution of the free propagator, $G_{0}(x',x,k)$.  The
Casimir energy depends on the boundary conditions obeyed by the
field $\psi$ and on the arrangement of the boundaries,
$\cS\equiv\partial \cD$ (not necessarily finite), of the domain
$\cD$.  From the outset we recognize that $\cE$ must be regulated,
and will in general be cutoff dependent, as discussed in the
Introduction.  We will not denote the cutoff dependence explicitly
except when necessary.  $\rho$ is the familiar density of states
associated with the problem
\begin{eqnarray}
    (\Delta+k^2)\psi(x)&=&0\mbox{\quad for\quad}x\in\cD\nonumber\\
    \label{eq:Cauchyprob}
    \psi(x)&=&0\mbox{\quad for\quad}x\in\cS,
\end{eqnarray}
so that $G$ satisfies the equation
\begin{eqnarray}
    \label{eq:Dirichletprob}
    (\Delta'+k^2)G(x',x,k)&=&-\delta^{N}(x'-x)\mbox{\quad for\quad}x', x
    \in\cD,
    \nonumber\\
    G(x',x,k)&=&0 \mbox{\quad for\quad} x' \ {\rm or}
     \ x \in \cS,
\end{eqnarray}
and
\begin{equation}
    \tilde G(x',x,k)=G(x',x,k)-G_{0}(x',x,k),
\end{equation}
where $G_{0}$ is the free scalar propagator in the absence of
boundaries.  The spectral representation expresses $G$ as a sum
over a complete set of eigenfunctions $\psi_n$ with eigenvalues
$k_n$
\begin{equation}
    \label{eq:spectraldec}
    G(x',x,k)=\sum_n\frac{\psi_n(x')\psi_n(x)}{k_{n}^2-k^2-i\epsilon}.
\end{equation}
Notice that since the problem (\ref{eq:Cauchyprob}) is real we
have chosen a complete set of \emph{real} eigenfunctions and
removed the usual complex conjugation from (\ref{eq:spectraldec}).
We can regard this problem as the study of a quantum mechanical
free particle with $\hbar=1$, mass $m=1/2$, and energy $E=k^{2}$,
living in the domain $\cD$ with Dirichlet boundary conditions on
$\partial \cD$. Dirichlet boundary conditions are an idealization
of interactions which prevent the quantum particle from
penetrating beyond the surfaces $\cS$.  This idealization is
adequate for low energies but fails for the divergent, {\it
i.e.\/} cutoff dependent, contributions to the Casimir
energy\cite{Graham:2003ib}.  As we have already seen in Section
II, the divergences can be simply disposed of in the optical
approach, and the physically measurable contributions to Casimir
effects are dominated by $k\sim 1/a$, where $a$, a typical plate
separation, will satisfy $1/a\ll \Lambda$ where $\Lambda$ is the
momentum cutoff characterizing the material.  So the boundary
condition idealization is quite adequate for our purposes.
Following this quantum mechanics analogy we introduce a fictitious
time, $t$, and consider the functional integral representation of
the propagator\cite{Feynman&Hibbs}.  The space-time propagator is
\begin{equation}
    G(x',x,t)=\int_{-\infty}^\infty \frac{dE}{2\pi
    i}G(x',x,\sqrt{E})e^{-iEt},
\end{equation}
where $E=k^{2}$.  Since $G$ is analytic in the upper half
$k^{2}$-plane, it is evident that $G(x',x,t)=0$ when $t<0$.  The
inverse Fourier transform reads
\begin{equation}
G(x',x,k)=i\int_0^\infty dt\ e^{ik^2t}G(x',x,t).
\label{kspace}
\end{equation}
$G(x',x,t)$ obeys the free Schr\"odinger equation in $\cD$ bounded
by $\cS$.  It can be written as a functional integral over paths
from $x'$ to $x$ with action $S(x',x,t)=\frac{1}{4}\int dt\dot
x^{2}$.
The optical approximation is obtained by taking the stationary
phase approximation of the propagator $G$ \emph{in the fictitious
time domain}.  Hence we assume that the functional integral is
dominated by the contribution of classical paths between $x'$ and
$x$.  These are straight line paths, reflecting $r$ times from the
boundaries, and traversed at constant speed, $v=\ell_{r}(x',x)/t$.
where $\ell_{r}(x',x)$ is the length of the path.  Then the
optical approximation to the propagator is given by,
\begin{equation}
\label{eq:tsemicl}
    G_{{\rm opt}}(x',x,t)=\sum_{r}D_r(x',x,t)e^{i
    S_r(x'x,t)}.
\end{equation}
The action is
\begin{equation}
    S_r(x',x,t)=\frac{\ell_r(x',x)^2}{4t}
\end{equation}
and $D$ is the van Vleck determinant
\begin{equation}
    D_r(x',x,t)\propto \det\left(\frac{\partial^2 \ell_r^2}{\partial
    x'_i\partial x_j}\right)^{1/2}
\end{equation}
This approximation is exact to the extent one can assume the
classical action of the path $S_r$ to be quadratic in $x',x$. This
is the case for flat and infinite plates.  Thus the non-quadratic
part of the classical action comes from the curvature or the
finite extent of the boundaries, which we parameterize generically
by $R$, $\partial^3 S/\partial x^3\sim 1/Rt$.  Hence, in a
stationary phase approximation $\delta x\sim \sqrt{t}$ and the
corrections are of order $\delta^3S\sim\Ord{\sqrt{t}/R}$.  Back in
$k$-space the corrections hence will be $\Ord{1/kR}$, and the
important values of $k$ for the Casimir energy are of order $1/a$,
where $a$ is a measure of the separation between the surfaces.
Thus the figure of merit for the optical approximation is $
{a/R}$.  At the moment there is no good way to estimate the order
in $a/R$ of the corrections to the optical approximation (possibly
fractional, plus exponentially small terms). Certainly some of the
curvature effects are captured by the van Vleck determinant, and
as we saw in section \ref{sec:sphereplane} for the sphere-plare
problem, the optical approximation works in practice out to
$a/R\sim 1$.  This is topic for further investigation.
Eq.~(\ref{eq:tsemicl}) is, in fact, the usual approximation of ray
optics, the van Vleck determinant being precisely the enlargement
factor of classical optics, as we now show.  Since $\partial
\ell_{r}(x',x)/\partial x' = \bfn'$ and $\partial
\ell_{r}(x',x)/\partial x =-\bfn $, where $\bfn$ and $\bfn'$ are
the unit tangent vectors to the path in the points $x$ and $x'$,
we have
\begin{equation}
         D_r(x',x,t)
        \propto\det\left(n_i n'_j+\ell_r\frac{\partial n_j}{\partial
        x'_i}\right)^{1/2},
\end{equation}
We perform the analysis in three dimensions.  Other values of $N$
are analogous, and we quote the general result at the end.  The
matrix $\frac{\partial n_j}{\partial x'_i}$ is
\begin{equation}
    \frac{d\phi_1}{dx'_1}
    \bft_1\otimes\bft'_{1}+
    \frac{d\phi_2}{dx'_2}
    \bft_2\otimes\bft'_{2},
\end{equation}
where $\bft_{1,2}$ and $\bft'_{1,2}$ are orthonormal tangent
vectors perpendicular to $\bfn$ and $\bfn'$ respectively and with
them form two orthonormal bases centered in $x$ and $x'$, and
$d\phi_i/dx'_i$ is the derivative of the angle subtended at the
point $x$ when we shift the point $x'$ along the direction
$\bft'_i$. Taking the determinant is now easy: it is the product
of the three eigenvalues of the matrix, but given the fact that
$\{\bfn,\bft_{1},\bft_{2}\}$ (and their primed correspondents) are
an orthonormal triple these are just $\{1,\ell_r
d\phi_1/dx'_1,\ell_r d\phi_2/dx'_2\}$, so
\begin{equation}
    D_r(x',x,t)\propto\left(\ell_r^2\frac{d\Omega_x}{dA_x'}\right)^{1/2}.
\end{equation}
The coefficient of proportionality is independent of the
path\footnote{We are not discussing the Maslov indexes other than the
$(-1)^r$ here.  If the ray $r$ would touch a caustic it would be
necessary to introduce the appropriate phase factor.} $r$ and must
depend on $t$ in such a way that for the direct path we obtain the
free propagator.  Therefore,
\begin{equation}
    D_r(x',x,t)=\frac{(-1)^{r}}{(4\pi i
    t)^{N/2}}\left(\ell_r^{N-1}\frac{d\Omega_x}{dA_x'}\right)^{1/2},
\end{equation}
where we have returned to $N$-dimensions.  We have introduced the
factor $(-1)^{r}$ to implement a Dirichlet boundary condition. In
the case of a Neumann boundary condition, this factor would not be
present.  Although we did not label $d\Omega/dA$ with an index
$r$, it should be clear from the derivation that it does depend on
the path $r$.
Putting all together we find the space-time form of the optical
propagator to be
\begin{equation}
    \label{eq:tdomain}
    G_{\rm opt}(x',x,t)=\sum_{ r}\frac{(-1)^{r}}{(4\pi i
    t)^{N/2}}\left(\ell_r^{N-1}\frac{d\Omega_x}
    {dA_x'}\right)^{1/2}e^{i\ell_r^2/4t}.
\end{equation}
When dealing with infinite, parallel, flat plates this
approximation becomes exact.  For a single infinite plate, for
example, the length-squared of the only two paths going from $x$
to $x'$ are
\begin{eqnarray}
    \ell_{{\rm direct}}^2&=&||x'-x||^2\nonumber\\
    \ell_{{\rm 1 reflection}}^2&=&||x'-\tilde{x}||^2,
\end{eqnarray}
where $\tilde x$ is the image of $x$.  Both are quadratic
functions of the points $x,x'$ and the optical approximation is
indeed exact.
In order to calculate the density of states we must return to
$k$-space.  $G(x',x,k)$ is obtained by Fourier transformation (see
eq.~(\ref{kspace})), and can be expressed in terms of Hankel
functions, giving us the final form for our approximation
\begin{eqnarray}
    G_ {\rm opt}(x',x,k)&=&\sum_r\frac{(-1)^{r}i\pi}{(4\pi
    )^{N/2}}\left(\ell_r^{N-1}\Delta_r
    \right)^{1/2}\left(\frac{\ell_r}{2k}\right)^{1-N/2}H^{(1)}_{\frac{N}{2}-1}
    \left(k\ell_r\right),\nonumber \\
    \label{eq:unifsemicl}
    &\equiv&\sum_r G_r(x',x,k),
\end{eqnarray}
where $N$ is the number of spatial dimensions, $\Delta_r$ is the
enlargement factor
\begin{equation}
    \label{eq:deltar}
    \Delta_r(x',x)=\frac{d\Omega_{x}}{dA_{x'}}
\end{equation}
and we have suppressed the arguments $x$ and $x'$ on $\ell_{r}$
and $\Delta_{r}$ in (\ref{eq:unifsemicl}).  This can be thought of
as a particular case of the general results in
Ref.~\cite{Berry72}.

For $N=1$ and $N=3$ the Hankel function reduces to an exponential.
For example, when $N=3$ we find
\begin{equation}
G_r(x',x,k)=(-1)^{n_r}\frac{\Delta_r^{1/2}}{4\pi}e^{ik\ell_r}.
\end{equation}
However, had we attempted a stationary phase approximation
directly in $k$-space we would have obtained an exponential
\emph{for any $N$},
\begin{eqnarray}
G_{{\rm semicl}}(x',x,t)=\sum_{{\rm paths\ }r}D_r(x',x,k)e^{i k
\ell_r(x'x)}\nonumber,
\end{eqnarray}
which does not reduce to the exact expression in the limit in
which we have only infinite, non-intersecting (hence parallel),
flat planes, because in $k$ space it is not a gaussian problem
even for quadratic $\ell^2$.  This is an important advantage of
applying the stationary phase approximation in the time domain
where it leads to the optical approximation.  Also we believe the
optical approximation to be a more favorable starting point for
considering systematic corrections to the stationary phase
approximation uniformly\footnote{The technique of passing to the
Fourier transform to obtain uniform approximations is certainly
not new in wave optics \cite{Kravtsov}.} in $1/R$.  The
expressions (\ref{eq:tsemicl}) and (\ref{eq:unifsemicl}) are the
first term of a systematic expansion of the propagator in $1/kR$.
For gently curved geometries we expect them to provide a good
approximation, the final test, in absence of exact solutions,
coming only from comparison with the experiments.  The corrections
come from two different (but related) effects \cite{Schulman}: a)
we have to expand the function $S_r(x',x)$ in the exponential to
include cubic (and higher order) terms and b) we have to include
other stationary paths of non-classical origin, like paths running
all around the bodies one or more times (these can be considered
as a non-perturbative, exponentially small correction to the
propagator). Both phenomena are due to the curvature of the
boundary surfaces and we go back to the previous estimate that the
parameter controlling the accuracy of the our approximation is
indeed $1/kR$ (wedges and discontinuities must be considered as
regions in which $R\to 0$ and the expansion is somewhat
different).  Two intertwined branches of wave optics have dealt
with finding corrections to the geometric optics predictions for
curved boundaries.  The first \cite{Keller,Kline} deals both with
perturbative \emph{a)} and nonperturbative \emph{b)} corrections
to next to leading order in $1/kR$ of particular importance in the
shadow region.  The second deals with edges and holes in locally
flat surfaces, originated by Sommerfeld's work \cite{BornWolf}
(see also \cite{Hannay} and references therein).  Both must be
considered relevant to future studies of Casimir forces, since
high-curvature and finite-size effects will soon be relevant in
the next generation of precision experiments
\cite{MPnew,MPprivate}.  Another phenomenon to be taken in
account, even in the case of gently curved surfaces, the optical
approximations fails when either $x$ or $x'$ are in the shadow
region or we are in presence of a caustic, the set of points where
the Hessian $\partial^2 S_r/\partial x\partial x'$ has one or more
vanishing eigenvalues \cite{PostonStewart}.  In these regions of
the parameters $(x,x')$ the gaussian approximation fails and one
cannot ignore cubic terms in the action.  There are various ways
of treating this phenomenon, whose importance in wave
optics\cite{Berry80, Kravtsov} as well as quantum mechanics
\cite{Berry72,Schulman} is today clear.  The most interesting
prediction related to the presence of caustics (for what concerns
us here) is the fact that a ray crossing a caustic acquires a
non-trivial phase shift.  This could possibly result in a change
of the sign of the Casimir force for concave geometry.
Unfortunately, the existing formalism does not seem to be easily
translated into our language and more work is needed in this
direction.

 The famous MRE of Balian and Bloch \cite{BalianBloch}
is also intimately related to the optical approximation developed
here.  It is relatively easy to see that our approximation arises
as the first term in a uniform $1/kR$ expansion for the
propagator. Most of the effort in applying the MRE to Casimir
energies has focused on the divergent terms associated with
general geometrical properties of the
bodies\cite{BalianDuplantier} or on the Casimir force at large
distances where only the lowest reflections contribute.  To our
knowledge no one has been able to develop a useful expansion
beyond the optical limit from the MRE.
\subsection{The optical Casimir energy}
The substitution of (\ref{eq:unifsemicl}) into (\ref{eq:rhoG}) and
then in (\ref{eq:casimiren}) gives rise to a series expansion of
the Casimir energy associated with classical closed (but not
necessarily periodic) paths
\begin{equation}
\cE_{\rm opt}=\sum_{{\rm paths\ } r} \cE_r,
\end{equation}
where each term of this series will be in the form of
\begin{equation}
\label{eq:rcontrib}
\cE_r=\frac{1}{2}\hbar\ \Im\int_0^\infty dk \omega(k)
\frac{2k}{\pi}\int_{\cD_r}d^Nx\ \ G_r(x,x,k).
\end{equation}
Here the integration has been restricted to the domain $\cD_r$
where the given classical path $r$ exists.
At this point it is useful to separate potentially divergent
contributions from those which are finite.  Because $G$ is
analytic in the upper half $k$-plane, the $k$ integration can be
taken along a contour with $\Im k>0$.  The Hankel function
$H^{(1)}_{n}(k\ell)$ falls exponentially in the upper half plane,
so the $x$ integral converges absolutely and uniformly at
fixed-$k$ unless there are $x$-values where $\ell_{r}(x)$ can
vanish.  One can easily convince oneself that for smooth
surfaces\footnote{It suffices that the vector {\bf n}
normal to the surface is continuous, i.e.\ no wedges are
present.} the only paths that reflect \emph{once} on any surface
can give vanishing path length. So for the moment, we put aside
the first reflection and consider the cutoff independent
contributions from $r>1$.  In that case we can interchange the $k$
and volume integrals.  The resulting $k$-integral is also
uniformly convergent.
\begin{equation}
    \cE_r=\frac{\pi\hbar}{2}\frac{(-1)^{r}}{(4\pi)^{N/2}}\ \Re\
    \int_{\cD_r}d^Nx\frac{(\ell_r^{N-1}\Delta_r)^{1/2}}{\ell_{r}^{N/2-1}}
    \int_0^\infty dk \omega(k) \frac{2k}{\pi}(2k)^{N/2-1}
    H^{(1)}_{N/2-1}(k\ell_r),
\end{equation}
for $r>1$.  The $k$-integral can be performed in general, but is
particularly simple for the massless case, $\omega(k)=ck$,
\begin{equation}
    \cE_r=\hbar
    c\frac{(-1)^{r+1}}{2\pi^{N/2+1/2}}\Gamma\left(
    \frac{N+1}{2}\right)
    \int_{\cD_r}d^Nx\frac{\Delta_r^{1/2}}{\ell_r^{(N+3)/2}},
    \label{ecasN}
\end{equation}
which is the Casimir energy associated to the optical path $r>1$,
and generalizes our fundamental result, eq.~(\ref{eq:cas1}) to
dimensions other than three. The generalization to the massive
case for $N=3$ is given by
\begin{equation}
    \cE_r=(-1)^{r+1}\frac{\hbar c\mu^{2}}{4\pi^{2}}
    \int_{\cD_r}d^3x\frac{\Delta_r^{1/2}}{\ell_r}K_2(\mu\ell_r),
    \label{ecasmass}
\end{equation}
which reduces to the $N=3$ case of eq.~(\ref{ecasN}) as $\mu\to
0$.  It is worth nothing that for $\mu>0$ the paths with
length $\ell\gtrsim 1/\mu$ are exponentially damped.
Now we return to analyze the potentially divergent first
reflection. For simplicity of notation we specialize to $N=3$
although the analysis is completely general. Let the boundary of
$\cD$ be the surfaces of a set of  rigid bodies $B_1,B_2,...,B_n$.
The divergent contributions come from the paths $1B_i$ that
reflect once on any of the bodies $B_i$.  To regulate possible
divergences we insert a simple exponential cutoff in $k$.  It is
easy to see that our results are independent of the form of the
cutoff.  Then for a massless field, reflecting from body $B$,
\begin{equation}
\cE_{1B}=(-1)\frac{\hbar c}{4\pi^{2}}\int_{\cD_{1B}}d^3x
\Delta_{1B}^{1/2}(x,x) \int_0^\infty dk\ e^{-k/\Lambda}k^2\
\sin(k\ell_{1B}(x,x)).
\label{1surf}
\end{equation}
The $k$-integration can be performed,
\begin{equation}
\cE_{1B}=-\frac{\hbar c}{4\pi^{2}} \int_{\cD_{1B}}d^3x\
\Delta_{1B}^{1/2}(x,x)
\frac{2\ell_{1B}\Lambda^4(3-(\ell_{1B}\Lambda)^2)}
{(1+(\ell_{1B}\Lambda)^2)^3}.
\label{1refldiv}
\end{equation}
Notice that for $\ell_{1B}\Lambda\gg 1$ we reobtain the standard
result, eq.~(\ref{eq:cas1}) as we should.  When
$\ell_{1B}\Lambda\lesssim 1$ however the structure of the function
changes completely. In particular the \emph{sign} changes at
$\ell_{1B}\Lambda=\sqrt{3}$. There is a non trivial
consequence of this fact: from eq.~(\ref{eq:cas1}) one expects a
positve divergence ($r=1$ here) as $\ell\to 0$, however the small
$\ell$ divergence in eq.~(\ref{1refldiv}) is negative. This
effect, that the cutoff dependent contribution to the Casimir
energy density changes sign near the bounding surface, is well
known and has figured centrally in recent discussions of Casimir
energy densities\cite{Graham:2002xq}.  Of course the bulk
contribution to the vacuum fluctuation energy comes from the
zero-reflection term, which is positive.  The negative surface
correction is well known and has many physical consequences.  For
example it contributes to the surface tension of heavy
nuclei\cite{Feshbach}.

To analyze the divergent first reflection, eq.~(\ref{1surf})
further, we need an expression for $\Delta(x,x)$ near a generally
curved surface.  This entails a small change in $\Delta_{1s}$ (see
eq.~(\ref{deltasphere})) to take in account two different
principal radii of curvature, say $R_a$ and $R_b$ (here $x'=x$ so
$\theta=0$ and $\sigma_1=\sigma_2=\ell/2$),
\begin{equation}
    \Delta_{1B}(x,x)=\frac{1}{(\ell_{1B}+\ell_{1B}^2/2R_a)
    (\ell_{1B}+\ell_{1B}^2/2R_b)}.
\end{equation}
Substituting back into eq.~(\ref{1refldiv}) and  replacing
$d^3x=dS(\ell)d\ell/2$, where
$dS(\ell)=(\ell/2R_a+1)(\ell/2R_b+1)dS$, and $dS$ is the surface
area element on the body, we get (up to finite terms arising from
upper bounds on the integration in $d\ell$)
\begin{equation}
    \cE\sim-\frac{\hbar c}{4\pi^{2}}\int dS\int_{0}^{\infty}d\ell
     \sqrt{(1+\ell/2R_a) (1+\ell/2R_b)}
    \frac{\Lambda^4(3-(\ell\Lambda)^2)}{(1+(\ell\Lambda)^2)^3}.
\end{equation}
where we have suppressed the subscript $1B$. The
$\ell$-integration may be evaluated at large $\Lambda$ to obtain
an asymptotic expansion of the cutoff dependent terms in the first
reflection,
\begin{eqnarray}
    \cE&\sim&-\frac{\hbar c}{4\pi^{2}}\int dS
    \left(\frac{\pi}{2}\Lambda^3+
    \frac{1}{8}\Lambda^2\left(\frac{1}{R_a}+\frac{1}{R_b}\right)
    +\Ord{\ln{\Lambda}}\right)\nonumber\\
    &=&-\frac{S}{8\pi}\hbar c\Lambda^3-\Lambda^2\frac{1}{32\pi^2}\hbar c\int
    dS\left(\frac{1}{R_a}+\frac{1}{R_b}\right)+\Ord{\ln
    \Lambda}
    \label{asymptdiv}
\end{eqnarray}
Eq.~(\ref{asymptdiv}) summarizes the cutoff dependent
contributions to the Casimir energy in the optical approximation.
As discussed in Section II, these terms do not contribute to the
forces between rigid objects.  Also they are trivial to isolate
and discard from the calculation of forces.
The form of eq.~(\ref{asymptdiv}) invites comparison with the work
of Balian and Bloch\cite{BalianBloch} on the asymptotic expansion
of the density of states based on their Multiple Reflection
Expansion.  The MRE propagator includes not only specular paths,
but also contributions from diffraction which also yield cutoff
dependent contributions to the Casimir Energy.  Scaling arguments
indicate that terms up to at least the third ``reflection''  in
the MRE are cutoff dependent. These higher divergences are omitted
from the optical approximation, which is convenient since they do
not contribute to Casimir forces in any case.  The first few terms
in the MRE expansion of the density of states are given by,
\begin{equation}
    \tilde\rho_{\rm MRE}(k)\sim
    2k\left(-\frac{S}{16\pi}-\frac{1}{12\pi^2k}\int
    dS\frac{1}{2}\left(\frac{1}{R_a}+\frac{1}{R_b}\right)+
    \Ord{1/k^2}\right),
\end{equation}
so the leading cutoff dependent terms in the Casimir energy
are\footnote{The sign of the second term here is opposite that of
Ref.~\cite{BalianBloch} because we are dealing with convex rather than
concave geometries}
\begin{eqnarray}
\cE &\sim& \frac{1}{2}\hbar c\int_0^\infty dk
 k\ \tilde\rho_{\rm MRE}(k)\ e^{-k/\Lambda}\nonumber\\
&\sim&-\frac{S}{8\pi}\hbar c\Lambda^3-\frac{1}{24\pi^2}\hbar
c\Lambda^2\int
dS\left(\frac{1}{R_a}+\frac{1}{R_b}\right)+\Ord{\Lambda}
\label{mrediv}
\end{eqnarray}
Comparing with the optical result, eq.~(\ref{asymptdiv}) we see
that the first terms agree and the second terms differ by a factor
of 3/4. Apparently our optical approximation to the propagator,
despite its simplicity, captures the leading divergence and the
order of magnitude of the subleading divergence\footnote{One might
think to claim more than order of magnitude success here. However
it should be noted that for Neumann boundary conditions both terms
in (\ref{asymptdiv}) change signs while only the surface terms in
(\ref{mrediv}) changes sign.  This is due to the fact that 2
`reflections' in the MRE expansion contribute to the curvature
divergence as well and their sign is the same for Dirichlet or
Neumann boundary conditions.}.
\subsection{Connections with other semiclassical approximations.}
\label{sec:connections}
Stationary phase approximations are not new in the study of
Casimir energy both at zero and non-zero temperature \cite{SandS,
Mazzitelli03}.  These works certainly share with ours the attempt
to switch the attention toward general properties and
approximations to the Helmholtz equation.  On the other hand,
relying more or less heavily on Gutzwiller's trace formula, they
suffer from two significant problems.  First, they treat symmetric
and nearly symmetric geometries in radically different ways, and
fail to provide a natural deformation away from the symmetric
limit (not to mention that they give the exact result for parallel
plates only for odd number of space dimensions). Second, they
require a certain amount of strongly geometry-dependent work (to
calculate monodromy matrices for example).  We discuss both these
problems further below. In order to study these points we will
rewrite a given contribution $\cE_r$ (specializing to $N=3$
dimensions and suppressing the index $r$) as
\begin{equation}
\label{eq:diverg1}
\cE=\Im(-1)^{n}\int_0^\infty dk\ \hbar c
k^2e^{-k/\Lambda}\int_0^\infty d\ell\ J(\ell)e^{ik\ell},
\end{equation}
where
\begin{equation}
J(\ell)\equiv\int_{\cD_r}d^3x\ \delta(\ell-\ell_r(x,x))\frac{
\Delta_r^{1/2}(x,x,k)}{4\pi^2}.
\end{equation}
Our strategy has been to dominate the functional integral over
paths from $x$ back to $x$ by the classical paths, then perform
the $k$ integral analytically, and to leave the integration over
$x$ for numerical evaluation.
The standard ``semiclassical'' approach \cite{Gutz,SandS} is to
perform all spatial integrations by stationary phase including the
one over the argument of the Greens function itself.  This leaves
a function only of $k$ which can be integrated analytically.  The
fact that we can do the $x$ integral numerically allows us to
capture much more detailed information about the system.  We will
show this in detail in the following.
To underline the differences, let us repeat briefly the line of
reasoning leading to Gutzwiller's trace formula. We start by
writing an asymptotic expansion for $k\ell\gg 1$. The asymptotic
contributions to the $ \ell$-integral come \cite{Bleistein} both
from a) boundaries at $\ell_m,\ell_M$ (minimum and maximum length
achieved by the path $r$) that is \emph{integration by part} terms
and b) integrable divergences in the function $J(\ell)$ that is
\emph{stationary phase} (SP) points at $\ell_j$. So that
\begin{equation}
\label{eq:Jexpansion}
J(\ell)\sim\sum_{n\geq
0}\left(A_n(k)e^{ik\ell_M}-B_n(k)e^{ik\ell_m}\right)+\sum_{{\rm
SP\ points\ }j }C_j(k)e^{ik\ell_j}.
\end{equation}
$A, B, C$ are polynomial in $k$, $1/k$ and
$\ell_{m},\ell_{M},\ell_{j}$ respectively. The Schaden and Spruch
\cite{SandS} approach based on Gutzwiller trace formula
\cite{Gutz} consists in taking only the stationary phase
contributions, b), to the energy, the coefficients $C_j(k)$'s then
being related to the `monodromy matrix'\footnote{To be precise the
stationary phase integral is done on the directions transverse to
the periodic orbit. The integration over the direction parallel to
the orbit is eventually performed by means of a trick
\cite{Gutz}.}. These terms correspond to closed classical paths,
for which the final momentum is equal to the initial one (the
action is $S\propto \ell$).  The stationary phase approximation
requires the periodic orbits to be well-separated in units of
wavelength.  However as one approaches a situation in which one
exact symmetry exists, the space, ${\mathbb R}^2$ in 3 dimensions,
perpendicular to the closed orbit at a given point breaks into the
product of two subspaces $A\otimes B$, and $\ell$ is constant with
respect to the $B$ coordinates $b$. In the symmetric situation the
SP\ points then form lines (or planes if more than one symmetry is
present) parameterized by $b$. The problem can again be solved
easily just by writing $d^2x\propto dadb$ and factoring out the
integral over $db$\cite{SandS,Creagh91} leaving the integral over
$da$ to be evaluated by stationary phase approximation
again\footnote{The simplest example is that of a cylinder facing a
plane. Then the periodic orbits are lines perpendicular both to
the cylinder and the plane, $b$ is parallel to the axis of the
cylinder and $a$ is the direction perpendicular to this. In the
case of parallel plates both the directions $a$ and $b$ are
symmetry directions so they both factor out and no stationary
phase approximation is performed. In this case the former analysis
gives an exact result, as is well known.}. However, when the
symmetry is slightly broken the length $\ell$ acquires a small $b$
dependence and the integral over $db$ can no longer be factored
out. Moreover a naive stationary phase approximation in both
$dadb$ is not reliable because arbitrarily close to the breaking
point, the dependence of $\ell$ on $b$ is small and the Hessian
matrix $\partial^2 \ell(x,x)/\partial x^2$ has one (or more) very
small eigenvalues in the old $b$ directions. There exists
\cite{Creagh96} a theory for Gutzwiller trace formula for
approximate symmetries. However, we found that it is not easy to
implement in the study of the Casimir energy for arbitrary
surfaces.
\begin{figure}[th]
%\centerline{ \BoxedEPSF{usandss.eps scaled 578}}
\centerline{\asfigure{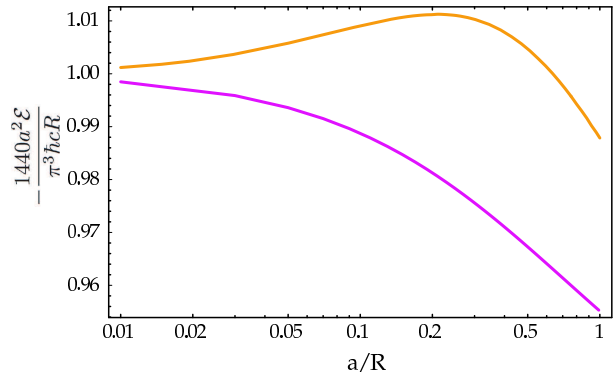}{578}{8cm}}
\caption{Comparison between the optical approximation (upper
curve) and the ``semiclassical'' approximation of Schaden and
Spruch (lower curve) for the sphere and plane.  The scaled Casimir
energy is plotted versus $a/R$. For the optical approximation, the
sum of the first four reflections has been rescaled to go to unity
as $a\to 0$. It is possible to show that in the limit $a/R\gg 1$
the optical approximation and Schaden and Spruch's formula agree
(in the figure they both tend to $90/\pi^4=0.92...$). The most
notable and relevant discrepancies are in the derivative at small
$a/R$.}
\label{fig:ratio}
\end{figure}
In the cases in which these problems can be avoided, like a sphere
of fixed radius in front of a plane, and for $a/R\ll 1$ the
semiclassical approximation \emph{\`a la} Schaden and Spruch
provides quite a good approximation (see figure) since in the
expansion (\ref{eq:Jexpansion}) the stationary phase approximation
gives a much larger contribution than the integration by parts
terms. A much stronger disagreement has to be expected if the
sphere gets substituted by a plate of width $w$ bent with a
curvature of order $R\gg w$.  Indeed the method of
Ref.~\cite{SandS} differs dramatically from the optical
approximation for the case of a hyperboloid\cite{SSJ}. This is not
a diffraction effect but rather a `precocious' breakdown of the
semiclassical approximation which is cured by a uniform
approximation of the kind we have described.
\section{Conclusions}
We have proposed a new method for calculating approximately
Casimir energies between conductors in generic geometries.  We use
a stationary phase approximation imported from studies of wave
optics that we have therefore named the ``optical approximation''.
In this paper, the first of the series, we have outlined the
derivation and applied it to three examples: the canonical example
of parallel plates; the experimentally relevant situation of a
sphere facing a plane; and the ``Casimir pendulum'', {\it i.e.\/}
a conducting plate free to oscillate above an infinite plate,
where the calculations can be performed analytically.  In all of
the above examples (except for parallel plates, where our result
coincides with Casimir solution) the agreement with the Proximity
Force Approximation is only to the leading order in the small
distances expansion.  The first order correction is found to be
different.  This is of particular importance in the example of the
sphere and the plane because the first order correction in $a/R$
($a$ is the distance sphere-plate and $R$ is the radius of the
sphere) will soon be measured by new precision experiments
\cite{MPprivate}. The optical approximation turns the Casimir sum
over modes into a sum over topologically different paths, and from
this point of view can be compared with the Poisson summation
formula, which has proved useful to derive semiclassical uniform
expansions for very diverse problems \cite{Berry72,Berry76}.  In
the case of the Casimir energy, replacing the usual highly
divergent sum over modes by a sum over topologically distinct
optical paths has two, very significant advantages: first, we have
been able to show that the divergences in the Casimir energy are
contained in contributions of very simple, one-reflection paths
and can be easily and unequivocally regulated and discarded; and
second, the convergence of the sum over paths is very rapid.
Instead of requiring an infinite number of eigenvalues with
exquisite precision one needs but a few path contributions,
calculated with little numerical effort, to give a very good
approximation to the Casimir energy for important geometries.
``Semiclassical'' methods have been used previously in the study
of Casimir effects and the connection between closed orbits and
the finite part of the Casimir energy has been pointed out many
times \cite{Fulling}. Our analysis shares with those the idea of
shifting attention to approximations and to properties of the
Helmholtz propagator.  We have shown however that in order to
obtain a correct low curvature approximation one has to use a
uniform approximation of the kind we proposed here. There is
plenty of room to improve the approximation presented here,
especially when the connection with Balian and Bloch's multiple
reflection expansion is made explicit. In particular it is
intriguing that the Casimir energy for the sphere-plane problem is
a well-defined problem in a single variable, namely $x=a/R$ whose
limiting values for $x\ll 1$ and $x\gg 1$ are famous
\cite{CasimirPolder,BalianDuplantier}. One can hope that an
analytic solution or an approximation good for the entire range
$x$-values should be relatively easy to find. On the contrary it
is an incredibly difficult problem and nobody has succeeded in
finding such an exact solution or a valid approximation. In the
next paper we will show how the same approximation for the
propagator can give useful expansions for local operators like the
energy-momentum tensor, which allows us to calculate the pressure
the energy density and other properties of the constrained field.
We will show what our approach has to say about the thermal
corrections, which are a controversial matter in Casimir physics.
In another paper of the series we will also perform the same
analysis for a field of spin 1/2 and and for the electromagnetic
field.
\section{Acknowledgements}
A.~S.\ would like to thank M.~V.~Berry and all the Theory Group of
H.~H.~Wills Laboratories at Bristol University, where part of this
work has been done, for their hospitality and for many
discussions. We would like to thank P.~Facchi, S.~Fulling,
I.~Klitch, L.~Levitov, L.~Schulman and F.~Wilczek, for early
discussions.  We are grateful to H.~Gies for correspondence and
numerical values of ref.\cite{Gies03}, to M.~Schaden and L.~Spruch
for correspondence and conversations regarding ref.\cite{SandS}
and to G.~Klimchitskaya and V.~Mostepanenko for discussions. This
work is supported in part by the U.S.~Department of Energy
(D.O.E.) under cooperative research agreement~\#DF-FC02-94ER40818.
A.~S.~is partially supported by INFN.
%%%%%%%%%%%%%%%%%%%%%%%%%%%%%%%%%%%%%%%%%%%%%%%%%%%%%%%%%%%%%%%%%

\end{document}